\documentstyle[12pt,fleqn]{article}
\topmargin - 0.8cm

\catcode`@=11
\@addtoreset{equation}{section}
\catcode`@=12
\mathchardef\SGamma="7100

\begin{document}
\title{\vskip-1.7cm \bf  Quantum Origin of the Early
Inflationary Universe}
\date{}
\author{A.O.Barvinsky$^{1}$, \ A.Yu.Kamenshchik$^{2}$
and \ I.V.Mishakov$^{2}$}
\maketitle
\hspace{-8mm}$^{1}${\em
Theory Department, Lebedev Physics Institute and Lebedev Research Center in
Physics, Leninsky Prospect 53,
Moscow 117924, Russia}
\\ $^{2}${\em Nuclear Safety
Institute, Russian Academy of Sciences , Bolshaya Tulskaya 52, Moscow
113191, Russia}
\begin{abstract}
We give a detailed presentation of a recently proposed mechanism
of generating the energy scale of inflation by loop effects in
quantum cosmology. We discuss the quantum origin of the early
inflationary Universe  from the no-boundary and tunneling
quantum states
and present a universal effective action
algorithm for the distribution function of chaotic inflationary
cosmologies in both of these states.
The energy scale of inflation is calculated by finding a
sharp probability peak in this distribution function for a
tunneling model driven by the inflaton field with large negative
constant $\xi$ of non-minimal interaction. The sub-Planckian
parameters of this peak (the mean value of the corresponding Hubble
constant $H\simeq 10^{-5}m_P$, its quantum width $\Delta H/H\simeq 10^{-5}$
and the number of inflationary
e-foldings ${\mbox{\boldmath $N$}}\geq 60$) are found
to be in good correspondence with the observational status of
inflation theory, provided the coupling constants of the theory
are constrained by a condition which is likely to be enforced by
the (quasi) supersymmetric nature of the sub-Planckian particle physics model.
\end{abstract}

\section{Introduction}
\hspace{\parindent}
In this paper we give a detailed account of a recently proposed
mechanism for generating the energy scale of the chaotic inflationary
Universe by the loop part of the effective action in quantum
cosmology \cite{Bar-Kam-Scale}.

Quantum cosmology became a theory of the quantum origin of inflationary
Universe in early eighties due to the synthesis of the cosmological
inflation \cite{Linde} with the idea of the quantum state
\cite{HH,H,Vilenkin}, generating the initial conditions for inflationary
scenario. One of the main problems of this theory was the formulation
of such a quantum state that could describe
a very early quantum Universe, its evolution leading to the
modern observable large-scale structure. The inflation paradigm is very
attractive because at least heuristically it
allows one to avoid applications of quantum gravity since the inflationary
epoch has to take place at the energy scale or a characteristic
value of the Hubble constant
$H=\dot a/a \sim 10^{-5}m_P$ much below the Planck one $m_P=G^{1/2}$.
The predictions of  the inflation theory essentially depend on this
energy scale which must be chosen to provide a sufficient number of
e-foldings ${\mbox{\boldmath $N$}}$ in the expansion law of a scale
factor $a(t)$ during the inflationary epoch, ${\mbox{\boldmath $N$}}=
\int _{t_{\rm I}}^{t_{\rm F}} dt\,H\geq 60$,
and also generate the necessary level of density perturbations.
This quantity, however, is a free parameter in the inflation theory,
and, to the best of our knowledge, there are no convincing principles that
could fix it without invoking the ideas of quantum gravity and cosmology.
These ideas imply that a quantum state of the Universe in the semiclassical
regime gives rise to an ensemble of inflationary universes with different
values of $H$, approximately evolving at later times according to
classical equations of motion. This quantum state
allows one to calculate the distribution function of this ensemble
and interpret its probability maximum at certain value of $H$ (if any) as
generating the quantum scale of inflation.

The implementation of this idea in the tree-level approximation of quantum
cosmology \cite{HH,H,Vilenkin,rhoT} has a controversial status and,
in our opinion, is not satisfactory. The corresponding distribution
functions are extremely flat \cite{H-Page,Vilenkin:tun-HH} for large values
of $H$ and unnormalizable at $H\rightarrow \infty$. This violates the
validity of the semiclassical expansion underlying the inflation
theory, since the contribution of over-Planckian energy scales
is not suppressed to zero, and special assumptions are necessary to
establish a Planckian "boundary" \cite{Linde-Mezh} to protect semiclassical
inflation physics from the nonperturbative realm of quantum gravity.
Apart from this difficulty, the possible local maxima of the distribution
function for the tree-level quantum states are either generating
insufficient amount of inflation violating the above bound \cite{GrisR},
or generate too high level of quantum inhomogeneities and require unnaturally
strong fine tuning (see Sect.2.1).

The key to the solution of these problems, not resorting to the conjectures
on a hypothetical over-Planckian phase of the theory, may consist in the
semiclassical  $\hbar$-expansion and the search for mechanisms that could
justify this expansion. Despite the perturbative nonrenormalizability of
quantum gravity, this approach makes sense in problems with quantum states
peaked at sub-Planckian energies. In particular, it would work in quantum
cosmology with the no-boundary \cite{HH,H} or tunnelling \cite{Vilenkin}
wavefunctions, provided they
suppress the contribution of Planckian energies and generate the probability
peaks at the lower (preferably GUT) scale compatible with microwave background
observations. As shown in authors' papers
\cite{BKam:norm,Bar-Rep,Bar-Kam-Scale,Bar-Kam-Tun}, the loop effects can
drastically change the predictions of the tree-level theory and
really allow one to reach this goal. Moreover, as it was briefly
announced in \cite{Bar-Kam-Scale}, one can
get a sharp probability peak in the distribution function of inflationary
models with characteristic parameters of GUT and, in this way,
provide a numerically sound link between quantum cosmology, inflation theory
and the particle physics of the early universe. Thus, the purpose of this
paper is to give a detailed presentation of this work.

The organization of the paper is as follows. Sec.2 presents quantum cosmology
as a theory of the quantum origin of the chaotic inflationary Universe. It
gives a brief account of the quantum gravitational tunnelling underlying the
no-boundary and tunnelling wavefunctions of the Universe, discusses the
model with nonminimally coupled inflaton field and presents a special
algorithm for the one-loop distribution function of (quasi)-DeSitter models.
Sects. 3 - 6 contain detailed calculations of various perturbative
contributions to this distribution function. We work within the double
perturbation theory: loop expansion in $\hbar$ up to the one-loop order and
the expansion of the slow roll approximation up to the subleading order in
the corresponding smallness parameter $m_P^2/|\xi|\varphi^2 \ll 1$, where
$\varphi$ is a value of the inflaton scalar field and $\xi=-|\xi|$ is
a big negative constant of its nonminimal coupling with curvature.
Important difference from our previous work \cite{Bar-Kam-Scale}, where the
distribution function was calculated only in the leading order of the
slow roll expansion, is that now we find it in a subleading order. This
leaves the main conclusions of \cite{Bar-Kam-Scale} qualitatively the same and
thus proves the stability of the leading order, although gives rise to
certain quantitative corrections. Sects. 5 and 6 contain perturbative
calculations of the one-loop effective action
for generic set of fields of various spins, contributing to the distribution
function. In Sect. 7 we present a final answer for this function and
find a corresponding probability peak that can be interpreted as generating
the energy scale of inflation. In Sect. 8 this result is used for the
derivation of the selection criterion for viable particle physics
models in the early Universe with nonminimal inflaton scalar field,
apparently suggesting their (quasi)supersymmetric nature. This conclusion is
based on the observation that the energy scale of inflation is suppressed
relative to the Planck scale by the same small factor $\sim 10^{-5}$
that determines a recently observed magnitude of the microwave background
radiation anisotropy \cite{COBE,Relikt}, provided that the two special
combinations of coupling constants of the system satisfy certain
restrictions. Sect.9 contains concluding remarks.

\section{Quantum cosmology -- the theory of quantum origin of
the early inflationary Universe}
\subsection{The no-boundary and tunneling quantum states}
\hspace{\parindent}
It is widely recognized now that the inflationary
scenario is one of the most promising pictures of the early Universe
\cite{Linde}. It can be described by the DeSitter or quasi-DeSitter
spacetime generated by an effective cosmological constant
$\Lambda$, which in its turn is being generated by other slowly varying
fields. Thus, in the model of chaotic inflation with the scalar
inflaton field $\phi$, minimally coupled to the metric tensor
$G_{\mu\nu}$
	\begin{eqnarray}
	L(G_{\mu\nu},\phi)=G^{1/2}\left\{\frac{m_P^2}
	{16\pi}R\,(G_{\mu\nu})-
	\frac 12 (\nabla\phi)^2-U(\phi)\right\},       \label{4.0}
	\end{eqnarray}
the effective cosmological constant is generated by the potential
of the inflaton field $U(\phi)$. For this system rolling down from the
potential barrier $U(\phi)$ (which is supposed to be monotonically
growing with $\phi$) there exists the so-called slow roll approximation,
when the non-stationarity of $\phi$ is much less than the rate of change
of the cosmological scale factor $a$ measured by the Hubble constant
$H=\dot{a}/a$. In this approximation equations of motion take the form
	\begin{eqnarray}
	&&\dot\phi\simeq -\frac1{3H}\frac{\partial U}
	{\partial\phi}\ll H\phi, \label{4.1} \\
        &&H=H(\phi)\simeq\sqrt{\frac{8\pi U(\phi)}{3m_P^2}},  \label{4.2}
	\end{eqnarray}
so that an effective cosmological constant $\Lambda=3H^2$ is determined
by the inflaton field potential. This potential is approximately
constant during the inflationary stage due to the slow change of
$\phi$ and only at the end of this stage decreases close to zero when
the effective cosmological constant "decays" into inflaton
oscillations, their energy being spent for the reheating of the Universe
and its transition to radiation-dominated and then
matter-dominated stages. The duration of the inflation stage usually
measured by the number of e-foldings  between the beginning
$t_I$ and the end $t_F$ of inflation
	\begin{eqnarray}
	{\mbox{\boldmath $N$}}=\int_{t_I}^{t_F} dt\,H
	\end{eqnarray}
determines the coefficient of inflationary expansion of the model
$\exp{\mbox{\boldmath $N$}}$
and depends on the initial value of the inflaton field $\phi_I$. This
dependence can be approximately obtained by changing the integration
variable here to $\phi$ and integrating from $\phi_I$ to zero. This
leads to a fundamental bound on the choice of inflationary model and
initial value of inflaton ${\mbox{\boldmath $N$}}\geq 60$ \cite{Linde}
	\begin{eqnarray}
	{\mbox{\boldmath $N$}}(\phi_I)\simeq \frac{4\pi}{m_P^2}
	\int_{0}^{\phi_I}d\phi\,\frac{H(\phi)}
        {\Big[\partial H(\phi)/\partial\phi\Big]} \geq 60.   \label{N}
	\end{eqnarray}

The role of quantum cosmology consists in the formulation of the quantum
initial data for such a picture in the form of a particular quantum state
of the Universe -- the wavefunction ${\mbox{\boldmath$\Psi$}}(q)$ usually
defined on superspace of 3-metric coefficients and all matter fields.
The implementation of this idea was proposed in the pioneering works of
Hartle, Hawking and Vilenkin
\cite{HH,H,Vilenkin}, who suggested that such initial
data (and the wavefunction ${\mbox{\boldmath$\Psi$}}(q)$) correspond
to the quantum gravitational tunneling that can semiclassically be
described by the transition with changing spacetime signature.

In the context of spatially closed cosmology the Lorentzian DeSitter
spacetime can be regarded as a result of quantum tunneling from
the classically forbidden state described by Euclidean DeSitter
geometry. A simple picture of the tunneling geometry demonstrating
such a mechanism is presented on Fig.1. The DeSitter solution of the
Einstein equations with the cosmological constant
$\Lambda=3H^2$
	\begin{eqnarray}
	&&ds^2_{L}=-dt^2+
	a^2_{L}(t)\,c_{ab}\,dx^{a}dx^{b},         \label{eqn:1.1}\\
	&&a_{L}(t)=
	\frac{1}{H}\,{\rm cosh}\,(Ht)            \label{eqn:1.2}
	\end{eqnarray}
describes the expansion of a spacelike spherical hypersurface with a metric of
a 3-sphere $a^2_{\!L}(t)\,c_{ab}$ with the radius (scale factor)
$a_L(t)$. Its Euclidean counterpart with the DeSitter positive signature
metric
	\begin{eqnarray}
	&&ds^2=d\tau^2+a^2(\tau)\,
	c_{ab}\,dx^{a}dx^{b},     \label{eqn:1.3}\\
	&&a\,(\tau)=
	\frac{1}{H}\,{\rm sin}\,(H\tau),               \label{eqn:1.4}
	\end{eqnarray}
represents a geometry of a 4-dimensional sphere of radius
$R=1/H$ with 3-dimensional sections (3-spheres) parameterized by a latitude
angle $\theta=H\tau$. These two metrics are related by the analytic
continuation into a complex plane of the Euclidean time
$\tau$ \cite{Mottola,Laflamme}
	\begin{eqnarray}
	\tau=\pi/2H+it,\;\;\;
        a_{L}(t)=a\,(\pi/2H+it).                     \label{eqn:1.5}
	\end{eqnarray}
This analytic continuation can be interpreted as a quantum nucleation
of the Lorentzian spacetime from the Euclidean one and shown on Fig.1
as a matching of the two manifolds (\ref{eqn:1.1}) - (\ref{eqn:1.4})
across the equatorial section of the 4-sphere $\tau=\pi/2H\;(t=0)$ --
the bounce surface $\Sigma_{B}$.

Two known quantum states which semiclassically implement this mechanism,
are represented by the no-boundary wave function
of Hartle and Hawking \cite{HH,H} and the tunneling wave function
\cite{rhoT} known for historical reasons as a wavefunction of Vilenkin
who pioneered the idea of quantum gravitational tunnelling in \cite{Vilenkin}.
In the approximation of the
two-dimensional minisuperspace consisting of the scale factor $a$ and
inflaton scalar field $\phi$
	\begin{eqnarray}
	q^i=(a,\,\phi),            \label{4.8}
	\end{eqnarray}
these wavefunctions ${\mbox{\boldmath$\Psi$}}_{N\!B}(a,\phi)$ and
${\mbox{\boldmath$\Psi$}}_{T}(a,\phi)$
satisfy the minisuperspace version of the Wheeler-DeWitt equation and
semiclassically represent its two linear independent solutions
	\begin{eqnarray}
	{\mbox{\boldmath$\Psi$}}_{N\!B}(a,\phi)\sim
	e^{\textstyle -{\mbox{\boldmath $I$}}(a,\phi)},\,\,\,
	{\mbox{\boldmath$\Psi$}}_{T}(a,\phi)\sim
	e^{\textstyle
	+{\mbox{\boldmath $I$}}(a,\phi)},               \label{4.9}
	\end{eqnarray}
where the Euclidean Hamilton-Jacobi function
${\mbox{\boldmath $I$}}(a,\phi)$ of the model is calculated at
a particular family of solutions of classical Euclidean equations
of motion subject to special boundary conditions of Hartle and Hawking at
$a=0$ (the "initial" point of the extremal) and boundary conditions
$(a,\phi)$ at the end point of the extremal --
an argument of the wavefunction. At $a=0$ the derivative of the scalar
field with respect to the Euclidean time $\tau$ should be zero while
$da/d\tau=1$ ($\tau$ measures the proper distance), which is
equivalent to the requirement of regularity of a 4-metric in the
neighbourhood of the pole of a 4-sphere (\ref{eqn:1.3}) -
(\ref{eqn:1.4}) at $\tau=0$. In the leading order of the
slow-roll approximation, when the inflaton field is constant, such a
solution coincides with the exact round 4-metric (\ref{eqn:1.3}) -
(\ref{eqn:1.4}) with the Hubble constant (\ref{4.2}), and its
Hamilton-Jacobi function equals
	\begin{eqnarray}
	{\mbox{\boldmath $I$}}(a,\phi)=
	-\frac{\pi m_P^2}{2H^2}
        \left[1-(1-H^2(\phi)\,a^2)^{3/2}\right],\,\,\,
	H^2(\phi)=\frac{8\pi
	U(\phi)}{3m_P^2}.                           \label{4.10}
	\end{eqnarray}

When the point $(a,\phi)$ belongs to the region of
minisuperspace below the curve (see Fig. 2)
	\begin{eqnarray}
        a=\frac 1{H(\phi)},              \label{4.11}
        \end{eqnarray}
the universe exists in the classically forbidden (underbarrier)
state described by this Euclidean spacetime. Euclidean extremals,
beginning at $a=0$, have a caustic
\footnote
{In the lowest order of the slow-roll approximation with
constant $\phi$ the problem is actually one-dimensional and Eq. (\ref{4.11})
represents a set of turning points, however beyond this
approximation this curve should, in fact, be replaced by the
envelope of the family of Euclidean trajectories
}
(\ref{4.11}) and cannot penetrate into the region $a>1/H(\phi)$ with
real Euclidean time. However, they can be continued into this region
by the analytic continuation into the complex
time (\ref{eqn:1.5}) which generates the Lorentzian (imaginary) part
of the Euclidean function ${\mbox{\boldmath $I$}}(a,\phi)$
	\begin{eqnarray}
	&&{\mbox{\boldmath $I$}}(a,\phi)=
	{\mbox{\boldmath $I$}}(\phi)\pm
	i{\mbox{\boldmath $S$}}(a,\phi),\,\,\,\,
	a>1/H(\phi), \label{4.12} \\
	&&{\mbox{\boldmath $S$}}(a,\phi)=
	-\frac{\pi m_P^2}{2H^2}
        \left(H^2\!(\phi)\,a^2-1\right)^{3/2}.  \label{4.13}
	\end{eqnarray}
Here ${\mbox{\boldmath $I$}}(\phi)$ is a Euclidean action of the
theory with the Lagrangian (\ref{4.0}) calculated on the
gravitational half-instanton -- the hemisphere (\ref{eqn:1.3})
-(\ref{eqn:1.4})
($0\leq\tau\leq\pi/2H$)
	\begin{eqnarray}
	{\mbox{\boldmath
	$I$}}(\phi)=-\frac{3m_P^4}{16U(\phi)}.   \label{4.14}
	\end{eqnarray}
This action determines the amplitude of wavefunctions (\ref{4.9}) in a
classically-allowed (Lorentzian) domain
	\begin{eqnarray}
	&&{\mbox{\boldmath$\Psi$}}_{N\!B}(a,\phi)\sim
	e^{\textstyle -{\mbox{\boldmath $I$}}(\phi)}
        \cos\left({\mbox{\boldmath $S$}}(a,\phi)
	+\pi/4\right),                              \label{4.15} \\
	&&{\mbox{\boldmath$\Psi$}}_{T}(a,\phi)\sim
	e^{\textstyle +{\mbox{\boldmath $I$}}(\phi)+
	i{\mbox{\boldmath $S$}}(a,\phi)},
	\,\,\,\,a>1/H(\phi),                         \label{4.16}
	\end{eqnarray}
which is interpreted in the tree-level approximation as
a distribution function for the one-parameter ensemble of Lorentzian
inflationary universes characterized by the Hamilton-Jacobi function
(\ref{4.13}). The parameter enumerating the members of this ensemble is
a value of the inflaton field $\phi$ or the corresponding
Hubble constant $H=H(\phi)$ and scalar curvature of the DeSitter
space. Its quantum distributions for the no-boundary
$\rho_{N\!B}(\phi)$ \cite{HH} and tunnelling $\rho_{T}(\phi)$ \cite{rhoT}
quantum states read
        \begin{eqnarray}
	\rho_{N\!B}(\phi)\sim e^
	{\textstyle -2 {\mbox{\boldmath $I$}}(\phi)},
	\,\,\, \rho_{T}(\phi)\sim
	e^{\textstyle +2
        {\mbox{\boldmath $I$}}(\phi)}.  \label{4.2.1}
	\end{eqnarray}

The difference between these two wave functions and their quantum
distributions consists in the different boundary
conditions in superspace: while the tunneling state
${\mbox{\boldmath$\Psi$}}_{T}(a,\phi)$ at $a>1/H(\phi)$ contains only
the outgoing wave and describes an expanding universe, the no-boundary
wave function ${\mbox{\boldmath$\Psi$}}_{N\!B}(a,\phi)$
in the Lorentzian regime represents the superposition of states
of expanding and contracting cosmologies corresponding to the
components of (\ref{4.15}) of positive and negative frequencies
with respect to the minisuperspace coordinate $a$. The tunneling wave
function is defined by the above mentioned outgoing wave conditions
in the Lorentzian region of superspace and an additional
condition of the $\phi$-independence of
${\mbox{\boldmath$\Psi$}}_{T}(a,\phi)$ at $a\rightarrow 0$
\cite{Vilenkin:tun-HH,VilVach}. For the no-boundary wavefunction there
exists a more fundamental and model-independent prescription in
the form of the path integral over regular Euclidean 4-geometries
\cite{HH,H}, which in the tree-level approximation is dominated by
the expression (\ref{4.15}) -- a contribution of the saddle point
of this integral -- the Euclidean-Lorentzian extremal (\ref{eqn:1.1}) -
(\ref{eqn:1.4}).

The distribution functions $\rho_{N\!B}(\phi)$ and  $\rho_{T}(\phi)$
describe the opposite results of the most probable underbarrier
tunneling: to the minimum and maximum of the inflaton potential
$U(\phi)\geq 0$ correspondingly (though in the latter case a minimum
$U(\phi)=0$ does not belong, strictly speaking, to the domain of
applicability of the slow-roll approximation).

Equations given above apply to the model (\ref{4.0}), however they can
also be used in the theory with non-minimally coupled scalar inflaton
$\varphi$
	\begin{eqnarray}
	{\mbox{\boldmath $L$}}(g_{\mu\nu},\varphi)
        =g^{1/2}\left\{\frac{m_P^2}{16\pi}R\,(g_{\mu\nu})-\!
        \frac12 \xi\varphi^2R\,(g_{\mu\nu})
        -\!\frac 12 (\nabla\varphi)^2
        -\!\frac12 m^2\varphi^2-\frac{\lambda}{4}\,
	\varphi^4\right\},                         \label{4.2.3}
        \end{eqnarray}
provided $L(G_{\mu\nu},\phi)$ above is viewed as the Einstein frame
of the Lagrangian ${\mbox{\boldmath $L$}}(\,g_{\mu\nu},\varphi\,)$
with the fields
$(G_{\mu\nu},\phi)=((1+8\pi|\xi|\varphi^2/m_P^2)g_{\mu\nu},\phi(\varphi))$
related to $(\,g_{\mu\nu},\varphi\,)$ by the known conformal
transformation of the metric and the reparametrization of the
scalar field \cite{SalopBB,Page:conf,BKK}.
For a negative nonminimal coupling constant $\xi=-|\xi|$ this model easily
generates the chaotic inflation scenario \cite{Spok-Unr} with the
following inflaton potential in the Einstein frame parameterization
	\begin{eqnarray}
        U(\phi)\,\Big|_{\,\textstyle\phi\!=\!\phi(\varphi)}=
        \frac{m^2\varphi^2/2+\lambda\varphi^4/4}
        {\Big(1+8\pi |\xi\,|\varphi^2/m_P^2\Big)^2} ,  \label{4.2.4}
        \end{eqnarray}
including the case of symmetry breaking at scale $v$ with
$m^2=-\lambda v^2<0$ in the Higgs potential
$\lambda(\varphi^2-v^2)^2/4$ . At large $\varphi$ the potential (\ref{4.2.4})
approaches a constant and depending on the parameter
	\begin{eqnarray}
	\delta\equiv -\frac{8\pi|\xi|m^2}{\lambda m_P^2}=
	\frac{8\pi|\xi|v^2}{m_P^2},                      \label{delta}
	\end{eqnarray}
has two types of behaviour at the intermediate values of the inflaton
field. For $\delta>-1$ it does not have local maxima and generates the
slow-roll decrease of the scalar field from its initial value $\varphi_I$
leading to a standard scenario with
a finite inflationary stage and approximate e-folding number
	\begin{eqnarray}
	{\mbox{\boldmath $N$}}(\varphi_I)=
	\left(\frac{\varphi_I}{m_P}\right)^2
	\frac{\pi(|\xi|+1/6)}{1+\delta}.               \label{N1}
	\end{eqnarray}
For $\delta<-1$ it has a local maximum at
$\bar\varphi=m/\sqrt{\lambda|1+\delta|}$, including the case of zero
$\lambda$ when $\bar\varphi=m_P/\sqrt{8\pi|\xi|}$, and due to
a negative slope of the potential leads to the inflation with
infinite duration for all models with the scalar field
growing from its initial value $\varphi_I>\bar\varphi$ to infinity.

The tree-level distribution functions (\ref{4.2.1}) for such a
potential do not suppress the over-Planckian
scales and are unnormalizable at large $\varphi$,
$\int^{\infty}d\varphi \,\rho_{N\!B,\,T}(\varphi) =\infty$,
thus invalidating a semiclassical expansion. Only for $\lambda=0$ the
normalizability takes place in the tunnelling case with $|\xi|\neq 0$,
$\rho_T(\varphi)\sim\exp[-48\pi^2|\xi|^2\varphi^2/m^2],\;
\varphi\rightarrow\infty$, but this fine tuning is too strong and can
hardly survive renormalization of $\lambda$ by quantum effects.

Only for a tunneling case with $\delta<-1$ the distribution
$\rho_{T}(\phi)$ has a local peak at the maximum of the potential
(\ref{4.2.4}) $\bar\varphi$, which could
serve as a source of the energy scale of inflation at reasonable
sub-Planckian value of the Hubble constant. However, this peak
requires a large positive mass of the
inflaton field $m^2>\lambda m_P^2/(8\pi|\xi|)$, which is too large
for reasonable values of $\xi=-2\times10^4$, $\lambda=0.05$
\cite{SalopBB}. Another (and maybe more serious) difficulty with this
inflation scenario starting from the maximum of inflation potential is
that according to (\ref{4.1}) it begins with zero
$\dot{\varphi}\sim\dot{\phi}=0$ and generates infinitely large quantum
inhomogeneities (inverse proportional to $\dot{\phi}$ \cite{Linde})
which are incompatible with the observable large scale structure of the
Universe. All this makes questionable the attempts to arrange the
quantum origin of our Universe at the tree-level theory and serves as
a motivation for considering the loop effects.

\subsection{One-loop distribution function of the inflationary cosmologies}
\hspace{\parindent}
Note that the calculation of the tree-level distribution
does not require the knowledge of the correct probabilistic inner
product of cosmological wave functions. It is enough to calculate and
square an amplitude of the wave function, which due to the
peculiarities of the model is a function on the section of the
two-dimensional minisuperspace, transversal to the coordinate $a$,
usually playing the role of time. Therefore the obtained
distribution function is defined on the physical subspace of a
correct dimensionality -- one-dimensional space of spatially homogeneous
inflaton field. Beyond the tree-level approximation the situation
changes: one needs the knowledge of the wave function with the
preexponential factor in the needed approximation, knowledge of the
correct inner product and the extension beyond the
minisuperspace approximation, because the distribution function
contains now a non-trivial contribution from integration over
inhomogeneous quantum fields frozen in the tree-level approximation. At the
one-loop level which we shall study here it is enough to consider
these fields in the linear approximation. Setting of the problem in the
model of chaotic inflationary Universe consists in the
minisuperspace model with the scale factor $a$ and spatially
homogeneous scalar inflaton $\varphi$ and with inhomogeneous fields of
all possible spins treated as perturbations on this background.
Together they form a superspace of variables
	\begin{eqnarray}
	q=(a,\,\varphi,\,\varphi({\bf x}),\, \psi({\bf
	x}),\,A_{a}({\bf x}),\, \psi_{a}({\bf x}),\,
	h_{ab}\,({\bf x}),...).                   \label{4.2.5}
        \end{eqnarray}
On this superspace we shall have
to calculate the no-boundary and tunnelling wavefunctions
${\mbox{\boldmath$\Psi$}}_{N\!B,T}(q)$ and then, by using a proper
physical inner product, calculate the distribution function of the
collective variable $\varphi$. To make the latter step and even to
give a rigorous definition of this distribution, it is better first
to make a quantum reduction to the wave function of physical ADM
variables $\xi$, which simultaneously disentangles time $t$ (initially
parametrized among the superspace variables (\ref{4.2.5}) \cite{ADM})
	\begin{eqnarray}
	q\rightarrow(\xi,t),\,\,\,
	{\mbox{\boldmath$\Psi$}}(q)\rightarrow\Psi(\xi,t).
	\end{eqnarray}

Strictly speaking this reduction is not consistent
(globally on phase space of the theory), and a complete
understanding and the interpretation of the cosmological
wavefunction might be reached only in the conceptually more advanced
framework (third quantization of gravity theory, refined
algebraic quantization \cite{Marolf,Landsman}, etc.). Although this framework still
does not have a status of a well-established physical theory,
there exists a good correspondence principle of this
framework with the quantization in reduced phase space for systems
with a wide class of special (positive-frequency) semiclassical
quantum states. For these states the conserved current of
the Wheeler-DeWitt equations perturbatively coincides with
the inner product of the ADM quantization and
thus can be used for the construction of the probability
distribution (for a perturbative equivalence of the ADM and
Dirac-Wheeler-DeWitt
quantization of gravity for such physical states see
\cite{BPon,BKr,Bar-Rep}). The tunneling
wavefunction belongs to such a class of states, while the no-boundary
one does not and should be supplied with additional (third quantization)
principles to be interpreted in terms of the probability distribution
of the above type.

It is plausible to make this reduction separately in the minisuperspace
sector of the full superspace $(a,\,\varphi)$ and its sector of
inhomogeneous modes. We choose an inflaton field $\varphi$ as a physical
variable whose distribution function we will calculate, while the
solution of classical equations of motion
(\ref{eqn:1.2}) with $H=H(\varphi)$ will be considered as a gauge
	\begin{eqnarray}
	\chi^{\perp}(a,\varphi,t)=a-\frac 1{H(\varphi)} {\rm
        cosh}(H(\varphi)t)=0.                    \label{4.2.6}
        \end{eqnarray}
It simultaneously plays the role of the parameterization of
minisuperspace coordinates in terms of the physical variable
\footnote
{This gauge is very convenient because it approximately
corresponds to the choice of the proper time with the lapse
function $N^\perp=1$ \cite{Bar-Rep}.
}.
The ADM reduction for linearized inhomogeneous modes of fields
boils down to the choice of their transverse
$(T)$ and transverse-traceless components
$(TT)$, so that the full set of physical variables reads
	\begin{eqnarray}
	\xi^A=(\varphi,f),\,\,\,f=(\varphi({\bf x}),\,
	\psi({\bf x}),\, A^{T}_{a}({\bf x}),\,
	\psi^{T}_{a}({\bf x}),\, h^{TT}_{ab}({\bf x}),...). \label{4.2.7}
	\end{eqnarray}

At the quantum level the ADM reduction can be easily carried out
by the method described in \cite{GenSem,Bar-Rep,BKr} for the tunneling state
(\ref{4.16}). However it stumbles upon the problem of positive and negative
frequency components for the Hartle-Hawking wavefunction
(\ref{4.15}) and in the gauge (\ref{4.2.6}) encounters the
analogue of the Gribov copies problem, corresponding to these
components. As discussed in \cite{AltBarv}, these copies are an
artifact of using inappropriate gauge, whose surface intersects
twice the classical extremal (\ref{eqn:1.2}) of one and the
same Universe before and after its bounce against the minimal value of a
cosmological radius $a=1/H(\varphi)$.  This implies a dubious interpretation
of (\ref{4.15}) as a superposition of two {\it simultaneously} existing
states of expanding and contracting Universe. This problem can be resolved
at the fundamental level by
the transition to quantization in conformal superspace in the
framework of the York gauge \cite{AltBarv}, but this framework is not yet
developed to be a workable technique. However, in the present model in the
semiclassical approximation it is enough merely to consider the
quantum ADM reduction for a separate positive frequency (or
negative frequency) component of (\ref{4.15}).

Thus, semiclassically for both cosmological states the quantum ADM reduction
boils down to obtaining the corresponding wave function of physical variables
$\Psi(\xi,t)=\Psi(\varphi,f|t)$. Then, the distribution function of
$\varphi$ should be regarded as a diagonal element
of the density matrix of this pure state ${\rm Tr}_f|\Psi\big>\big<\Psi|$.
It can be obtained from $|\Psi\big>=\Psi(\phi,f\,|\,t)$ by averaging
over the rest of the modes of physical fields $f$
	\begin{eqnarray}
        \rho\,(\phi\,|\,t)= \int
        df\;\Psi^*(\phi,f\,|\,t)\, \Psi(\phi,f\,|\,t),  \label{4.3.1}
	\end{eqnarray}
and does not reduce to a simple squaring of the wave function.

The calculation of the one-loop no-boundary and tunnelling wavefunctions
perturbatively in inhomogeneous modes $f$ on the Friedmann-Robertson-Walker
background was carried out in many papers
\cite{Fisch,HHal,Laflamme,Wada,VilVach,BKam:norm,Bar-Kam-Tun,Bar-Kam-Scale}.
It can be based on the path integration over the fields
regular on the Euclidean spacetime with metric
(\ref{eqn:1.3}) - (\ref{eqn:1.4}) or by using the known one-loop
approximation for the general solution of the Wheeler-DeWitt equations
\cite{GenSem,Bar-Rep,BKr}. Then both wavefunctions turn out to be Gaussian
in the variables $f$ -- their Euclidean DeSitter invariant vacuum
\cite{Laflamme}. Therefore the integration over $f$ in (\ref{4.3.1}) is
trivial and leads to the fundamental algorithm which is valid for both
no-boundary \cite{BKam:norm,Bar-Rep,Bar-Kam-Tun,reduct,Bar-Kam-Scale}
and tunneling \cite{QGrav} quantum states
	\begin{eqnarray}
        \rho_{N\!B,T}(\varphi|t)\cong
         \frac{\Delta_\varphi^{1/2}}{|v_\varphi(t)|}\; e^
	{\textstyle\, \mp 2{\mbox{\boldmath $I$}}(\varphi)-
        {\mbox{\boldmath $\Gamma$}}
	_{\rm 1-loop}(\varphi)}.                       \label{4.3.2}
        \end{eqnarray}

It involves the doubled Euclidean action on the hemisphere with the
metric (\ref{eqn:1.3}) - (\ref{eqn:1.4}), the linearized mode of the
homogeneous inflaton field $v_\varphi(t)$ -- the basis function of the
wave equation on the Lorentzian DeSitter background (\ref{eqn:1.1}) -
(\ref{eqn:1.2}), which can be obtained from the {\it regular} Euclidean
linearized mode $u_\varphi=u_\varphi(\tau)$
	\begin{eqnarray}
	\frac{\delta^2 I[\,\xi\,]}
	{\delta \xi\,\delta \xi}\, u_\varphi=0      \label{4.3.2a}
	\end{eqnarray}
by the analytic continuation (\ref{eqn:1.5})
	\begin{eqnarray}
	&&\frac{\delta^2 S[\,\xi\,]}
	{\delta \xi\,\delta \xi}\, v_\varphi=0,       \label{4.3.2b}\\
	&&v_\varphi(t)=
	\left[u_\varphi(\pi/2H+it)\right]^*.      \label{4.3.2c}
	\end{eqnarray}
Here $I[\,\xi\,]$ and $S[\,\xi\,]$ are the Euclidean and
Lorentzian action functional reduced to the ADM physical variables, so that
$\delta^2 I[\xi]/\delta\xi\,\delta\xi$ and
$\delta^2 S[\xi]/\delta\xi\,\delta\xi$ are the operators of their physical
inverse propagators. The mode $v_\varphi(t)$ is normalized in (\ref{4.3.2})
to unity with respect to the
Wronskian inner product in the space of solutions of this wave equation
	\begin{eqnarray}
	\Delta_\varphi=iv_\varphi^*(t)
	\stackrel{\leftrightarrow}{W}
	v_\varphi(t),                                   \label{4.3.3}
	\end{eqnarray}
where $\stackrel{\leftrightarrow}{W}=\stackrel{\rightarrow}{W}-
\stackrel{\leftarrow}{W}$ is a corresponding Wronskian operator linear in
time derivative.
The one-loop contribution is the same for both quantum states and is
determined by the Euclidean effective action of all physical fields $\xi(x)$
	\begin{eqnarray}
	{\mbox{\boldmath $\Gamma$}}_{\rm 1-loop}(\varphi)=
        \left.\frac12\,{\rm Tr\,ln}\,
        \frac{\delta^2 I[\,\xi\,]}{\delta \xi\,\delta \xi}
        \,\right |_{\,\rm \bf D\!S}.                    \label{4.3.4}
	\end{eqnarray}
This effective action is calculated on the DeSitter instanton
-- 4-dimensional sphere of the radius $1/H(\varphi)$ --
and, therefore, is a function of $\varphi$ -- the
argument of the distribution function. Such a closed Euclidean
manifold is obtained by the doubling of the half-instanton
\cite{Bar-Kam-Tun} -- two hemispheres match each other along the
equatorial hypersurface $\Sigma_B$ (on which a quantum transition
with the change of signature takes place). Graphical illustration of
the procedure of calculating the distribution function is given on
Fig. 3. The wave function and its conjugate, participating in the scalar
product (\ref{4.3.1}), can be represented by two Euclidean-Lorentzian
manifolds. When calculating the inner product, due to implicit
unitarity of the theory, the contributions of Lorentzian regions cancel
each other and the result boils down to the Euclidean effective action
calculated on a closed instanton obtained by gluing the two
hemispheres of the above type \cite{Bar-Kam-Tun,reduct}.

The above simple scheme holds only in the lowest order of the
slow-roll approximation, when the
inflaton scalar field is constant on the solution of classical
(Euclidean and
Lorentzian) equations of motion and generates the effective Hubble
and cosmological constants invariable in time. This is the case of the
so-called real tunnelling geometry \cite{Hal-Hartle}. In the chaotic
inflation model, however, the inflaton field is varying with time which
makes the spacetime geometry deviating from the exact DeSitter one, and
correspondingly the analytic continuation from the Euclidean regime to the
Lorentzian one makes the extremal complex. The general case of complex
gravitational tunneling was considered in much detail in \cite{Bar-Kam-Tun}
and for this particular model in \cite{Lyons}.
The modification due to the imaginary part of the complex extremal looks
as follows. The complex minisuperspace extremal $Q(z)=(a(z),\varphi(z))$
in the complex plane of the Euclidean time $z=\tau+it$ should satisfy
the following boundary-value problem
	\begin{eqnarray}
	&&\frac{\delta {\mbox{\boldmath $I$}}[Q]}
	{\delta Q(z)}=0,\,\,\,
	\left.\frac{d\varphi(z)}{dz}\right|_{z=0}=0,\,\,
	a(z)\sim z+O(z^2),\,z\rightarrow 0,         \label{4.3.5}\\
	&&Q(z_+)=q_+=(a_+,\varphi_+)                 \label{4.3.6}
	\end{eqnarray}
with the Hartle-Hawking no-boundary conditions at $z=0$ and real boundary
data $(a_+,\varphi_+)$ at the boundary of the spacetime ball $z=z_+$. The
Euclidean action ${\mbox{\boldmath $I$}}(a_+,\varphi_+)=
{\mbox{\boldmath $I$}}[Q(z)]$ calculated on the solution
of this problem is in general
complex but cannot be decomposed as before in the purely real
contribution of
the Euclidean part of the full manifold and imaginary contribution of
its real Lorentzian section. Thus, the algorithm (\ref{4.3.2})
still holds,
but with ${\mbox{\boldmath $I$}}(\varphi)$ replaced by the real part of the
complex action
	\begin{eqnarray}
	{\mbox{\boldmath $I$}}(\varphi)\rightarrow
	\Re{\rm e}\,{\mbox{\boldmath $I$}}[Q(z)]=
	\Re{\rm e}\,{\mbox{\boldmath $I$}}(a_+,\varphi_+)  \label{4.3.7}
	\end{eqnarray}
and with all the other quantities calculated on the background of
this extremal \cite{Bar-Kam-Tun}. The final real point of the extremal
$(a_+,\varphi_+)$ should be a subject of the ADM reduction which identifies
$\varphi_+$ with the physical variable and expresses $a_+$ as a function of
the physical time $t_+$ and $\varphi_+$ in some gauge, like (\ref{4.2.6}). In
what follows we shall find such a complex extremal in the first subleading
order of the slow-roll approximation and calculate the corresponding
distribution function. The model we consider will be a chaotic inflationary
cosmology with the inflaton field non-minimally coupled to curvature with
a large negative coupling constant $\xi=-|\xi|$. As we
shall see, a small parameter of the slow roll expansion in this model
turns out to be inverse proportional to
$|\xi|$, $m_P^2/|\xi|\varphi^2\ll 1$. This choice is
justified by the fact that this model with $|\xi|\simeq 2\times 10^4$ is
regarded as a good candidate for the inflationary scenario compatible with
the observational status of inflation theory \cite{SalopBB}.

\section{Non-minimal inflaton scalar field: tree-level approximation}
\subsection{Perturbation theory for Euclidean classical solutions}
\hspace{\parindent}
Let us find the classical solution of the problem (\ref{4.3.5}) in the model
(\ref{4.2.3}) with the nonminimally coupled inflaton field. We shall
begin solving this problem on the Euclidean segment of complex time
$z=\tau$. The Euclidean action of this model
in the minisuperspace of homogeneous scale factor $a$, inflaton
$\varphi$ and lapse $N$ looks like
         \begin{eqnarray}
         &&{\mbox{\boldmath$I$}}[N,a,\varphi] =-2 \pi^{2}
        \int_{\tau_{-}}^{\tau_{+}} d \tau N a^{3} \left\{\frac{1}{2}
        (\frac{3m_{P}^{2}}{4 \pi}-6 \xi
        \varphi^{2})\left(\frac{1}{a^{2}}-
        \frac{\dot{a}^{2}}{a^{2}}\right)
	-\frac{1}{2} \dot{\varphi}^{2} \right. \nonumber\\
	&&\qquad\qquad\qquad\qquad\qquad\qquad\qquad\qquad
	\left.- 6 \xi \varphi \dot{\varphi}
	\frac{\dot{a}}{a}-\frac{1}{2} m^{2} \varphi^{2} -
        \frac{1}{4}\lambda\varphi^{4}\right\}.    \label{miniaction}
	\end{eqnarray}
where the dot denotes the differentiation with respect to proper
Euclidean time $d/Nd\tau$.

The first order variational derivatives of this action
with respect to $\varphi$ and $N$ give the Euclidean
equations of motion
	\begin{eqnarray}
        &&{\mbox{\boldmath$I$}},_{\varphi}\equiv -
	A \left\{\ddot{\varphi}
	+ 3 \frac{\dot{a}}{a} \dot
        {\varphi}-\left[m^{2} +
	\lambda \varphi^{2}
	+ 6\xi\left(\frac{1}{a^{2}} -
	\frac{\dot{a}^{2}}{a^{2}} -
	\frac{\ddot{a}}{a}\right)\right]
	\varphi\right\}=0,                          \label{variscal}\\
        &&{\mbox{\boldmath$I$}},_{N}
         \equiv -\frac{A}{N} \left\{\frac{1}{2}
	\left(\frac{3m_{P}^{2}}{4 \pi} - 6\xi\varphi^{2}\right)
        \left(\frac{1}{a^{2}} - \frac{\dot{a}^{2}}{a^{2}}\right) +
        \frac{1}{2} \dot{\varphi}^{2} + 6\xi\varphi \dot{\varphi}
        \frac{\dot{a}}{a}\right. \nonumber\\
	&&\qquad\qquad\qquad\qquad\qquad\qquad\qquad\qquad\qquad
	\left.- \frac{1}{2} m^{2} \varphi^{2}
        -\frac{1}
        {4} \lambda \varphi^{4}\right\}=0,         \label{varilapse}
	\end{eqnarray}
where the normalization factor $A= 2 \pi^{2} N a^{3}$.
Eq. (\ref{variscal}) is a dynamical equation of motion for the
scalar field $\varphi$, while Eq. (\ref{varilapse}) is a
Lagrangian version of the Hamiltonian constraint. There is no need to
write down the variation of the action with respect to the scale factor
$a$ because this equation is a consequence of (\ref{variscal}) -
(\ref{varilapse}).

We now develop the perturbation theory for the above equations
starting with the DeSitter solution as a lowest order approximation of a
constant scalar field $\varphi$. This solution in the cosmic time gauge
$N=1$ has the form
	\begin{eqnarray}
	\varphi^{(0)} = \varphi_{0}={\rm const},
	\;\;a^{(0)} = \frac{1}{H(\varphi_0)}
	\sin [H(\varphi_0)\,\tau],               \label{DeSitter}
	\end{eqnarray}
where the bracketed superscript denotes the order of perturbation theory
and $\varphi_0=\varphi(0)$ is the initial value of the inflaton scalar field
at $\tau=0$ and $a=0$. Substituting it into Eq. (\ref{varilapse}) we
obtain $H(\varphi)$ as a following function of the scalar field:
	\begin{eqnarray}
        H^{2}(\varphi) = \frac{m^{2} \varphi^{2} +
        \lambda \varphi^4/2}{3m_{P}^{2}/4\pi
	-6\xi\varphi^{2}}=
        \frac{\lambda \varphi^2}{12|\xi|}
	\left[\,1 - \frac{ m_P^2(1+2\delta)}
	{8\pi|\xi|\varphi^{2}}+O\,\left(\frac {m_P^4}
	{|\xi|^2\varphi^4}\right)\,\right].         \label{Hubble}
	\end{eqnarray}
On the solution (\ref{DeSitter}) the equation (\ref{varilapse}) is satisfied
exactly, while the equation (\ref{variscal}) holds only approximately
	\begin{eqnarray}
	&&{\mbox{\boldmath$I$}},_{\varphi }^{(0)} =
	A_{0} M_{0}^{2} \varphi_{0},        \label{scalonDS}\\
	&&M_{0}^{2}\equiv M^{2}(\varphi_0)=
	\left.\left[m^{2} + \lambda\varphi^{2}
	+12\xi H^{2}(\varphi)\right]
	\right|_{\varphi=\varphi_{0}}         \label{effectmass}
	\end{eqnarray}
with the effective mass in the dynamical equation for inflaton field
$M^2(\varphi)$ which is small in view of (\ref{Hubble}) (remember that
$\xi=-|\xi|<0$) in the limit of large $|\xi|$ and $\varphi^2$
	\begin{eqnarray}
	M^{2}(\varphi) = \lambda m^{2}_P
	\frac{1+\delta}{8\pi|\xi|}+
	O\,(m_P^4/|\xi|^2\varphi^2),          \label{effectmass1}
	\end{eqnarray}
provided the parameter $\delta$ defined by Eq.(\ref{delta}) is bounded
(in the chaotic inflation model with nonminimal inflaton the constant
$|\xi|$ is usually chosen of the order of magnitude of ratio of the
Planck scale to the GUT scale $v=-m^2/\lambda$, so that $\delta\ll 1$).
This property of cancellation of the leading in $\varphi$ contributions to
(\ref{effectmass}) underlies the slow-roll approximation in this model
of the nonminimal and nonlinear inflaton field. This approximation works
for large values of the inflaton, its inverse playing the role of
smallness parameter. As we shall now see, big negative $\xi$ further
improves this expansion which actually takes place in powers of
$m_P^2/|\xi|\varphi^2$.

To find the first subleading order of this expansion, we have to
expand the equations of motion (\ref{variscal})-(\ref{varilapse})
up to the first order in perturbations
	\begin{eqnarray}
	&&{\mbox{\boldmath$I$}},_{N \varphi} \delta\varphi +
	{\mbox{\boldmath$I$}},_{N a} \delta a =
	-{\mbox{\boldmath$I$}},_{N}\equiv 0,    \label{linconstraint}\\
	&&{\mbox{\boldmath$I$}},_{\varphi\varphi}
	\delta \varphi +
	{\mbox{\boldmath$I$}},_{\varphi a} \delta a
	= - {\mbox{\boldmath$I$}},_{\varphi}^{(0)}   \label{secvarscal3}
	\end{eqnarray}
and solve this linear system for $\delta \varphi$ and $\delta a$. Here
we use an obvious notations for the second order variational derivatives
of the action, which are the differential operators evaluated
at the lowest-order solution (\ref{DeSitter}). With the choice of a new
(angular) variable on the DeSitter sphere, replacing the Euclidean time,
	\begin{eqnarray}
	\theta = H(\varphi_0)\,\tau  \label{theta}
	\end{eqnarray}
these operators take the form
	\begin{eqnarray}
	&&{\mbox{\boldmath$I$}},_{\varphi\varphi} =
	-A H^{2}\left\{\frac{d^{2}}{d\theta^{2}}
	+ 3 \cot \theta \frac{d}{d\theta}
	-\left(\frac{m^{2}}{H^{2}}
	- 24\xi\right)\right\},             \label{secvarscal2}\\
	&&{\mbox{\boldmath$I$}},_{\varphi a} =
	-A H^{2} \frac{6\xi\varphi_{0}}{a}
	\left\{\frac{d^{2}}{d\theta^{2}}
	+ 2 \cot \theta \frac{d}{d\theta} + 3\right\}
	+\frac{3}{a}
	{\mbox{\boldmath$I$}},_{\varphi},  \label{secvarscalcosm}\\
	&&{\mbox{\boldmath$I$}},_{N \varphi} =
	A H^{2}\left\{6\xi\varphi_{0}
	\left(1-\cot \theta \frac{d}{d\theta}\right)
	+\frac{m^{2}\varphi_{0} + \lambda
	\varphi_{0}^{3}}{H^{2}}\right\},    \label{secvarlapsescal}\\
	&&{\mbox{\boldmath$I$}},_{N a} =
	\frac{A H^{2}}{a} (3m_{P}^{2}/4\pi
	-6\xi\varphi_{0}^{2})\left(1 +
	\cot \theta \frac{d}{d\theta}\right),\label{secvarlapsecosm}
	\end{eqnarray}
with the normalization factor $A$, Hubble constant $H$ and scale factor $a$
taken in the lowest-order approximation.

For large $\varphi_0$ and in view of (\ref{Hubble}) the last two
equations can be simplified
	\begin{eqnarray}
	&&\frac{1}{A H^{2}}
	{\mbox{\boldmath$I$}},_{N \varphi}\delta\varphi
	= - 6\xi \varphi_{0} \frac{\cos^{2}\theta}{\sin\theta}
	\frac{d}{d \theta}
	\frac{\delta \varphi}{\cos\theta}
	+ O(m_P/\varphi_{0}), \label{secvarlapsescal1}\\
	&&\frac{1}{A H^{2}}
	{\mbox{\boldmath$I$}},_{N a}\delta a
	= - 6\xi \varphi_{0}^{2} H
	\frac{\cos^{2}\theta}{\sin^{2} \theta} \frac{d}{d \theta}
	\frac{\delta a}{\cos\theta}
	\left(1+O(m_P^2/\varphi^2_0)\right)    \label{secvarlapsecosm1}
	\end{eqnarray}
and used in the linearized constraint equation (\ref{linconstraint})
to give
	\begin{eqnarray}
	\frac{d}{d \theta} \frac{\delta a}{\cos\theta} =
	- \frac{\sin\theta}{\varphi_{0} H_{0}} \frac{d}{d \theta}
	\frac{\delta \varphi}{\cos\theta}
	+O(m_P^5/\varphi^5_0).          \label{linconstraint1}
	\end{eqnarray}

This equation in its turn can be used in the dynamical linearized
equation (\ref{secvarscal3}). The latter in view of
(\ref{secvarscal2})-(\ref{secvarscalcosm}) can be rewritten in
the form
	\begin{eqnarray}
	&&\left\{\frac{d^{2}}{d\theta^{2}}
	+ 3 \cot \theta \frac{d}{d\theta}
	+ 24 \xi\right\} \delta \varphi +
	6\xi\varphi_{0}H \left\{
	\frac{d}{d\theta} + 3 \cot \theta - \tan \theta
	\right\}\frac{\cos\theta}{\sin\theta}\frac d{d\theta}
	\frac {\delta a}{\cos\theta} \nonumber \\
	&&\qquad\qquad\qquad\qquad\qquad\qquad
	=\frac{M_{0}^{2}\varphi_{0}}{H^{2}}
	+O(m_P^3/\varphi^3_0),                  \label{secvarscal4}
	\end{eqnarray}
which allows one to exclude $\delta a$ on account of (\ref{linconstraint1})
and thus arrive at the closed equation for $\delta\varphi$
	\begin{eqnarray}
	\left\{\frac{d^{2}}{d\theta^{2}} + 3 \cot \theta
	\frac{d}{d\theta}\right\} \delta \varphi = \frac{1}{1-6\xi}
	\frac{M_{0}^{2} \varphi_{0}}{H^{2}}+O(m_P^3/\varphi^3_0).
	\label{eqofmotion}
	\end{eqnarray}
Its solution satisfying the Hartle-Hawking
boundary conditions \cite{HH,H} $(d/d\theta)\delta\varphi(0) = 0,
\ \ \delta\varphi(0) = 0$, reads (we use (\ref{Hubble}) and
(\ref{effectmass1})
	\begin{eqnarray}
	\delta \varphi = - \frac{m_P^2}{\pi\varphi_0}
	\frac{1+\delta}{1-6\xi}
	\left(\ln \cos \frac{\theta}{2}
	- \frac{1}{4}\tan^{2}\frac{\theta}{2}
	\right) + O(m_P^3/\varphi_{0}^3).      \label{deltavarphi}
	\end{eqnarray}
The corresponding perturbation of the scale factor $\delta a$ can then be
obtained by integrating the equation (\ref{linconstraint1}) with zero initial
condition at $\theta=0$
	\begin{eqnarray}
	&&\delta a = \frac{m_P^2}{\pi\varphi_0^2
	H(\varphi_0)} \frac{1+\delta}{1-6\xi}
	\cos\theta\left(\tan\theta
	\ln \cos\frac{\theta}{2}
	- \frac{1}{4}\tan^{2}\frac{\theta}{2}
	\tan\theta \right.\nonumber \\
	&&\qquad\qquad\qquad\qquad\left.-\frac{\theta}{4}
	+ \frac{1}{2}\tan\frac{\theta}{2}
	-\int_{0}^{\theta} d\theta'
	\ln \cos \frac{\theta'}{2}\right)
	+O(m_P^5/\varphi_{0}^5).                     \label{deltacosm}
	\end{eqnarray}

The results of this section can be summarized in the one-parameter family
of Euclidean extremals enumerated by the initial value of the inflaton
field $\varphi_0$
	\begin{eqnarray}
	&&\varphi(\tau,\varphi_0)=\varphi_0+
	\delta\varphi(\theta,\varphi_0), \label{varphi}\\
	&&a(\tau,\varphi_0)=\frac 1{H(\varphi_0)}\sin\theta
	+\delta a(\theta,\varphi_0),\,\,\,
	\theta\equiv H(\varphi_0)\tau  \label{a}
	\end{eqnarray}
with perturbations $\delta\varphi=O(m_P/\varphi_0)$ and $\delta a=
O(m_p^3/\varphi_0^3)$ given above. These perturbations are small for large
$\varphi_0$. The corresponding smallness parameter is $m_P^2/\varphi_0^2$
(one should remember that for a bounded range of $\theta$ and
in the leading order $\varphi\sim\varphi_0,\,
a\sim \varphi_0^{-1}$). A direct inspection of these perturbations also
shows that for large negative constant of nonminimal coupling $|\xi|\gg
1$ the actual smallness parameter is $m_P^2/|\xi|\varphi_0^2$, and the
final conclusions of this paper will be obtained  for
$m_P^2/|\xi|\varphi_0^2 \ll 1$.

\subsection{Superspace caustic and complex extremals}
\hspace{\parindent}
Once we have a perturbative solution parametrized by the initial
value of the scalar field $\varphi_0$, we
can now perturbatively solve the boundary value problem (\ref{4.3.6})
which takes the form of two equations
	\begin{eqnarray}
	&&\varphi_0+\delta\varphi(\theta_+,\varphi_0)=
	\varphi_+,                            \label{3.2.1}\\
	&&\frac 1{H(\varphi_0)}\sin\theta_+
	+\delta a(\theta_+,\varphi_0)=a_+,     \label{3.2.2}
	\end{eqnarray}
for $(\varphi_0,\,\theta_+)$ in terms of $(\varphi_+,\,a_+)$. Like
(\ref{varphi})-(\ref{a}) this solution can be obtained in a
subleading approximation in the form
	\begin{eqnarray}
	\varphi_0=\varphi_0^{(0)}+\varphi_0^{(1)},\,\,
	\theta_+=\theta_+^{(0)}+\theta_+^{(1)},    \label{3.2.3}
	\end{eqnarray}
where the superscript in brackets denotes the order of perturbation theory.

The lowest order approximation
	\begin{eqnarray}
	&&\varphi_0^{(0)}=\varphi_+,\\
	&&\theta_+^{(0)}=\arcsin\left[H(\varphi_+)\,a_+\right]
	\end{eqnarray}
shows that only the points $(\varphi_+,a_+)$ lying below the curve
	\begin{eqnarray}
	a = \frac{1}{H(\varphi)}   \label{caustics}
	\end{eqnarray}
can be reached by Euclidean trajectories (with the no-boundary initial
conditions) in real time. This curve consists of the points of
maximal expansion of the Euclidean model at $\theta=\pi/2$ and can be
regarded as a caustic of the family of solutions in the lowest order
of the slow roll expansion. The points of two-dimensional
minisuperspace above this curve can be reached only in complex Euclidean
time because
	\begin{eqnarray}
	&&\theta_+^{(0)} = \frac{\pi}{2} +
	i H(\varphi_+) t_{+},           \label{Lorentztime}\\
	&&H t_{+} = {\rm arccosh}
	\left[H(\varphi_{+}) a_{+}\right],         \label{thetaplus}
	\end{eqnarray}
where $t_+$ can be interpreted as the Lorentzian time in the DeSitter
space nucleating at $t_+=0$ from the Euclidean DeSitter hemisphere
parameterized by the angular coordinate $\theta=H(\varphi_{+})\tau,\,
0\leq\theta\leq \pi/2$. Thus in this case the combined Euclidean-Lorentzian
evolution takes place on a contour in a complex plane of time,
consisting of the Euclidean $0\leq\tau\leq \pi/2H(\varphi_{+})$ and
Lorentzian $\tau=\pi/2H(\varphi_{+})+it,\,0\leq t \leq t_+$ segments.

In the lowest order of the slow roll approximation the point of
Euclidean-Lorentzian transition coincides with the point of maximal
expansion of $a$ on the Euclidean trajectory where
the both velocities $\dot{\varphi}$ and $\dot{a}$ vanish. This situation
corresponds to the so-called real tunnelling \cite{Hal-Hartle} when the
analytically matched Euclidean and Lorentzian extremals both have real values
of the configuration space coordinates. Beyond the leading order of the
slow roll approximation this property does not hold.

The first subleading approximation for $(\varphi_0,\,\theta_+)$ can be
obtained by making in Eqs. (\ref{3.2.1})-(\ref{3.2.2}) the first-order
iteration in $\delta\varphi$ and $\delta a$
	\begin{eqnarray}
	&&\varphi_{0}^{(1)} =
	-\delta\varphi(\theta_{+}^{(0)},
	\varphi_+),                           \label{varphi^{1}}\\
	&&\theta_{+}^{(1)} =
	-\frac 1{\cos \theta_+^{(0)}}
	\left (H(\varphi_+)\,\delta a(\theta_{+}^{(0)},\varphi_+)+
	\sin \theta_+^{(0)}\,\delta\varphi(\theta_{+}^{(0)},\varphi_+)
	/\varphi_+ \right )=                    \nonumber\\
	&&\qquad\qquad\frac{m_P^2}{\pi\varphi_0^2}
	\frac{1+\delta}{1-6\xi}
	\left(\frac{\theta_{+}^{(0)}}{4} -
	\frac{1}{2}\tan\frac{\theta_{+}^{(0)}}{2}
	+\int_{0}^{\theta_{+}^{(0)}} d\theta
	\ln \cos \frac{\theta}{2}\right)
	+O(m_P^4/\varphi_{+}^4).                       \label{theta1}
	\end{eqnarray}
In this approximation the point of maximal expansion
(extremum of (\ref{3.2.2})) at $\theta_m$ differs from $\pi/2$
	\begin{eqnarray}
	&&\theta_m = \frac{\pi}{2} + \varepsilon_{m}, \\
	&&\varepsilon_{m} = \frac{m_P^2}{\pi\varphi_0^2}
	\frac{1+\delta}{1-6\xi} \left(-\frac{3}{2} +
	\frac{\pi}{8}
	-\frac{\pi}{2}\ln 2 + G \right),      \label{maxicosm2}
	\end{eqnarray}
where $G$ is the Catalan constant ($G =\int_0^{\pi/4} d\theta
\ln\cot\theta=0.915 \ldots$). The set of these points, however,
does not form the caustic curve: in contrast with the lowest order
of the slow roll approximation, in which the dynamics is actually
one-dimensional, $\dot{\varphi}_0=0$, now the both
minisuperspace coordinates depend on time and the equation that
determines the envelope of the $\varphi_0$-parameter family of
extremals (\ref{varphi})-(\ref{a}) is given by the following requirement.
At the envelope curve the vector tangential to the extremal
$(\dot{\varphi}, \dot{a})$ and the vector transversal to their one
parameter family $(\partial \varphi/\partial \varphi_{0},
\partial a/\partial \varphi_{0})$ are collinear, i.e.
	\begin{eqnarray}
	\dot{\varphi} \frac{\partial a}{\partial \varphi_{0}}
	-\dot{a}\frac{\partial \varphi}
	{\partial \varphi_{0}} = 0.      \label{causticsequation}
	\end{eqnarray}
From this equation it follows that the extremal starting at $\tau=0$ with
the initial scalar field $\varphi_0$ hits the caustic at
	\begin{eqnarray}
	&&\theta_c(\varphi_0) = \frac{\pi}{2} +
	\varepsilon_c(\varphi_0), \label{caustics1}\\
	&&\varepsilon_{c}(\varphi_0) =
	\frac{m_P^2}{\pi\varphi_0^2}
	\frac{1+\delta}{1-6\xi}  \left(-\frac{1}{2} +
	\frac{\pi}{8} -\frac{\pi}{2}
	\ln 2 + G \right).        \label{caustics2}
	\end{eqnarray}
The caustic curve when parametrized by the value of $\varphi_0$ can be
obtained by substituting $\theta_c(\varphi_0)$ in (\ref{varphi}) and
(\ref{a}). Then, with $\varphi_0$ excluded in terms of $\varphi$ from the
second of the resulting equations, the first of them becomes
the equation of the caustic as a graph of $a$ against $\varphi$. Simple
calculations show that the first order corrections cancel out and the
caustic equation remains the same as the leading-order one (\ref{caustics})
$a = 1/H(\varphi)+O(m_P^4/\varphi^5)$. The meaning of this curve implied by
the equation (\ref{causticsequation}) is that its points belonging to
a particular extremal $a=a(\tau,\varphi_0),\,\,\varphi=
\varphi(\tau,\varphi_0)$ are no longer in one-to-one correspondence
with the value of the Euclidean time $\tau$ and initial value of the
scalar field $\varphi_0$ enumerating the extremals
	\begin{eqnarray}
	\frac{\partial\,(\varphi,a)}{\partial\,(\tau,\varphi_0)}=0.
	\end{eqnarray}
One cannot go beyond this curve in real Euclidean time: the continuation
in real $\theta>\theta_c(\varphi_0)$ results in the trajectory bouncing
back to the Euclidean domain under the caustic curve
$a<1/H(\varphi)+O(m_P^4/\varphi^5)$. Similarly to the lowest order
approximation
the points beyond this curve can be reached only in complex time. Now,
however, we have two important differences. Firstly, the both velocities
$(\dot{a},\,\dot{\varphi})$ at the nucleation point (as anywhere else on
every extremal) are nonvanishing simultaneously which results in the
complex valuedness of the extremal ending at the point $(\varphi_+,a_+)$
beyond the caustic (the discussion of matching conditions at the
Euclidean-Lorentzian transition for this case of complex tunnelling can
be found in the previous authors' work \cite{Bar-Kam-Tun}). And, secondly, the point of
nucleation (the corresponding value of the Euclidean time) is a nontrivial
function of the end point of the extremal. To show this take
$a_{+} > 1/H(\varphi_+)$ and calculate the final value of the complex
Euclidean time $\theta_+$ given by Eqs.(\ref{3.2.3}),
(\ref{Lorentztime}), (\ref{thetaplus}), (\ref{theta1}). It has
a real part
	\begin{eqnarray}
	&&\Re {\rm e}\,\theta_+ \equiv\theta_N(\varphi_+,t_+) =
	\frac{\pi}{2} + \varepsilon_{N}(\varphi_+,t_+), \\
	&&\varepsilon_{N}(\varphi_+,t_+)=\frac{m_P^2}{\pi\varphi_+^2}
	\frac{1+\delta}{1-6\xi}
	\left(-\frac{1}{2 \cosh\,[\,H(\varphi_+\!)\,t_{+}]}
	+ \frac{\pi}{8}-\frac{\pi}{2} \ln 2 + G \right.  \nonumber\\
	&&\qquad\qquad\qquad\qquad\qquad\left.+\frac12 H(\varphi_+\!)
	\int_{0}^{t_{+}} dt
	\arctan\sinh\,[\,H(\varphi_+\!)\,t\,] \right),   \label{nucleation1}
	\end{eqnarray}
which can be identified with the nucleation point provided we choose
the contour of complex ``angular'' time $\theta$ joining the points
$\theta=0$ and $\theta_+$ with the union of two segments $0\leq \theta
\leq \theta_N$ and $\theta=\theta_N+iH(\varphi_+)\,t,\,
\,0\leq t\leq t_+$. In contrast with the case of real tunnelling this
is just a convention, because on both segments the fields are complex and
analyticity does not give any preference to this particular choice.
Comparison of $\varepsilon_m,\,\varepsilon_c$ and $\varepsilon_N$ shows that all these
deviations from the lowest order nucleation point $\theta=\pi/2$ are
different: $\varepsilon_m<\varepsilon_c$ and $\varepsilon_N=\varepsilon_c$
only at $t_+=0$, that is when the end point of the extremal lies on the
caustic, while for positive Lorentzian time $\varepsilon_N>\varepsilon_c$
\footnote
{Note that our perturbation theory for $\theta_N$ and other Lorentzian
quantities in its present form is valid only for small values of the
Lorentzian time $t,\,\,H(\varphi_+)\,t_+\sim 1,$ to be valid in the slow roll
limit of big $H(\varphi_+)$. In the Euclidean context one of the motivations
for introducing the coordinate $\theta=H\tau$ instead of $\tau$
was the fact the range of $\theta$ below the caustic is bounded by $\pi/2$
which is different from the Lorentzian domain where the hyperbolic
``angle'' $Ht$ is unbounded from above.
}.

Once we have a solution for equations
(\ref{4.3.5})-(\ref{4.3.6}) we can calculate the corresponding
Hamilton-Jacobi function and, in particular, its real part
(\ref{4.3.7}) $\Re{\rm e}\,{\mbox{\boldmath $I$}}(a_+,\varphi_+)$. When
calculated in the lowest order approximation on the solution (\ref{DeSitter})
the doubled real part of the action
obviously coincides with the Euclidean action on the full four-dimensional
sphere of the radius $1/H(\varphi_+)$ and constant scalar field $\varphi
=\varphi_+$. When expanded in powers of $m_P^2/\varphi^2_+$ it equals
	\begin{eqnarray}
	&&2\Re{\rm e}\,{\mbox{\boldmath $I$}}(a_+,\varphi_+)
	= I_{0} +  \frac{I_{1}}{\varphi_+^{2}}
	+ O\left(\frac {m_P^4}{\varphi_+^4}\right),   \label{actionexpansion}\\
	&&I_{0} = -\frac{96 \pi^{2}
	|\xi|^{2}}{\lambda},                   \label{action0}\\
	&&I_{1} = -\frac{24 \pi m_{P}^{2}
	|\xi| (1 + \delta)}{\lambda}.       \label{action1}
	\end{eqnarray}
Since we want to have this quantity in the first subleading approximation
in $m_P^2/\varphi^2_+$, a priori we have to include corrections to this result
linear in $\delta\varphi$ and $\delta a$. By direct calculations these
corrections can be shown to vanish due to certain intrinsic cancellations.
The mechanism of these cancellations follows from the parametrized nature
of the gravity theory and looks as follows. Using the notations of
(\ref{4.3.5})-(\ref{4.3.6}) (with complex $\theta$ replacing $z$) we can
write down the total action on the extremal subject to boundary conditions
$q_+$ at $\theta_+$ as an integral over $\theta$ of the corresponding
Lagrangian
	\begin{eqnarray}
	{\mbox{\boldmath $I$}}_{\theta_+}[\,Q(\theta)\,]
	=\int_{0}^{\theta_{+}} d\theta \,L(Q,\dot{Q})   \label{actionL}
	\end{eqnarray}
with a subscript indicating the upper limit of integration $\theta_+$
over the complex angular ``time''which is a solution of the boundary
condition (\ref{4.3.6}). We have a solution of classical equations of
motion as
a perturbation expansion $Q(\theta,\varphi_0)=Q^{(0)}(\theta,\varphi_0)
+\delta Q(\theta,\varphi_0)$ and in two subsequent orders of this
perturbation theory the boundary conditions have the form
	\begin{eqnarray}
	&&Q^{0}(\theta^{(0)}_{+},\varphi_0^{(0)})=q_{+}\\
	&&Q^{(0)}(\theta^{(0)}_{+}+\theta^{(1)}_{+},\varphi^{(0)}_0
	+\varphi^{(1)}_0)+
	\delta Q(\theta^{(0)}_{+},\varphi^{(0)}_{0})=q_{+},
	\end{eqnarray}
whence it follows that
	\begin{eqnarray}
	\left[\,\frac{\partial Q^{(0)}}{\partial\varphi_{0}}
	\varphi^{(1)}_{0}+
	\dot{Q}^{(0)}\theta^{(1)}_{+}
	+\delta Q\,\right]_{\theta_+}=0,  \label{relation}
	\end{eqnarray}
where the dot denotes the derivative with respect to $\theta$.
Now, the total action (\ref{actionL}) calculated in a first
subleading order of perturbation theory reads
	\begin{eqnarray}
	&&{\mbox{\boldmath $I$}}_{\theta_{+}^{(0)}+\theta^{(1)}_{+}}
	\left[Q^{(0)}(\theta,\varphi_{0}^{(0)}
	+\varphi^{(1)}_{0})+\delta Q\right]  \nonumber\\
	&&\qquad\qquad\qquad
	={\mbox{\boldmath $I$}}_
	{\theta^{(0)}_{+}}\left[\,Q^{(0)}(\varphi^{(0)}_{0},\theta)\,\right]
	+L\,\theta^{(1)}_{+}+
	\left[\,\frac{\partial L}{\partial \dot{Q}}\left(\frac{\partial
	Q^{(0)}}{\partial\varphi_{0}}\varphi^{(1)}_{0}
	+ \delta Q\right)\,\right]_{\theta_+}
	\end{eqnarray}
(here the first order variation of the functional argument of the action
reduces to the standard surface term at $\theta_+$, because this
variation is being calculated at the solution of the classical equations).
Then from (\ref{relation}) the total first order correction induced by
$\delta Q$ reduces to
	\begin{eqnarray}
	\left(L-\frac{\partial L}{\partial \dot{Q}}\dot{Q}\right)
	\theta^{(1)}_{+}=0
	\end{eqnarray}
and vanishes because the coefficient of $\theta^{(1)}_{+}$ here in a
reparametrization invariant theory boils down to a Hamiltonian constraint
identically satisfied for a classical background. Thus the final expression
for the real part of the classical action reduces in the subleading
approximation to (\ref{actionexpansion})-(\ref{action1}).

The absence of the first-order corrections in $\Re{\rm e}\,{\mbox{\boldmath $I$}}
(a_+,\varphi_+)$ due to $\delta Q$ guarantees the tree level unitarity of
the theory -- the time independence of the semiclassical wavefunction
amplitude. As it follows from(\ref{actionexpansion}) it depends only on the
final value of the scalar field $\varphi_+$, while the contribution of
$\delta Q$ could have introduced a nontrivial dependence on $a_+$ (or $t_+$).
This property, obvious for real tunnelling, for complex extremals was
shown to be a consequence of the Einstein-Hamilton-Jacobi equation for
complex semiclassical phase \cite{Hal-Hartle,Bar-Kam-Tun}.

\section{The homogeneous inflaton mode}
\hspace{\parindent}
Here we calculate the contribution of the Lorentzian inflaton mode
$v_{\varphi}(t)$ to the probability distribution $\rho(\varphi, t)$,
given by the preexponential factor $\Delta_\varphi^{1/2}/|v_\varphi(t)|$
of (\ref{4.3.2}). The algorithm (\ref{4.3.2}) was obtained within the
ADM reduced phase-space quantization implying that $\varphi$ is a
physical degree of freedom while the scale factor $a$ and the lapse function
$N$ are determined by the gauge condition (\ref{4.2.6}). This gauge condition
introduces cosmic time (with unit lapse $N=1$) \cite{Bar-Rep},
therefore the Euclidean and Lorentzian wave equations for $v_{\varphi}(t)$
(\ref{4.3.2a})-(\ref{4.3.2b}) can be obtained by the same linearization
procedure as in Sect.3 (performed in $N=1$ gauge).
After exclusion of $\delta a$ of this section, this equation
boils down to the homogeneous version of Eq.(\ref{eqofmotion})
with $u_{\varphi}(\tau)$ replacing $\delta\varphi$
	\begin{eqnarray}
	\frac{\delta^2 I[\,\xi\,]}
	{\delta \xi\,\delta \xi}\, u_\varphi=
	\frac{2\pi^2}{H} \left\{-\frac{d}{d\theta}\sin^3\theta
	\frac{d}{d\theta}
	+ O(m_P^2/|\xi|\varphi_0^2)\right\}
	u_\varphi = 0                   \label{eqinflmode}
	\end{eqnarray}
(one should remember that the operator in (\ref{eqofmotion})
enters the quadratic part of the action with the factor
$A\,H^2=2\pi^2 a^3H^2\approx 2\pi^2 \sin^3\theta/H$). This equation was,
however, obtained only in the lowest order approximation
while we would need to know $v_{\varphi}(t)$ in a subleading order of the
slow roll expansion. The remedy is to use the one-parameter family of
classical solutions already known in this approximation. The needed
linearized mode can be obtained by differentiating this solution (\ref{a})
with respect to $\varphi_0$
	\begin{eqnarray}
	u_{\varphi}(\tau) = \frac{\partial
	\varphi(\tau,\varphi_{0})}
	{\partial \varphi_{0}}                   \label{homogen}
	\end{eqnarray}
and then analytically continuing it to the Lorentzian spacetime
(\ref{4.3.2c}). Using (\ref{deltavarphi}) one has the Euclidean mode
in the first subleading approximation
	\begin{eqnarray}
	u_{\varphi}(\tau) = 1 + \frac{m_P^2}{2\pi\varphi_0^2}
	\frac{1+\delta}{1-6 \xi}
	\left\{\theta\,\frac{2+\cos\theta}{1+\cos\theta}
	\tan\frac{\theta}{2}+ 2 \ln \cos \frac{\theta}{2}
	-\frac{1}{2}\tan^{2}\frac{\theta}{2}\right\}.   \label{homogen1}
	\end{eqnarray}
Since it has the form $1+O(m_P^2/|\xi|\varphi_0^2)$ for $|\xi|\gg 1$
the modulus of the corresponding Lorentzian mode gives the contribution to the
exponential of $\rho(\varphi, t)$
	\begin{eqnarray}
	\ln \frac 1{|v_{\varphi}(t)|} =
	O(m_P^2/|\xi|\varphi_0^2)          \label{homogencorrection}
	\end{eqnarray}
which is by one power of $1/\xi$ smaller in magnitude
than the subleading term $I_{1}/\varphi_{0}^{2}$ of the corresponding
tree-level contribution (\ref{actionexpansion}).

Now we have to calculate the Wronskian norm of this mode
$\Delta_{\varphi}$. From (\ref{eqinflmode}) it follows that the
Wronskian operator in the Euclidean time $\tau=\theta/H$
($\stackrel{\leftrightarrow}{W}{}^{\!\!E}=
\stackrel{\rightarrow}{W}{}^{\!\!E}-
\stackrel{\leftarrow}{W}{}^{\!\!E}=i\stackrel{\leftrightarrow}{W}$) equals
	\begin{eqnarray}
	\stackrel{\rightarrow}{W}{}^{\!\!E}\equiv
	W^E = -2\pi^2 \frac{1}{H^2}
	\sin^{3} \theta \frac{d}{d\theta}.
	\end{eqnarray}
The time-independent
inner product (\ref{4.3.3}) can be calculated at the nucleation point $t=0$
where it equals $2u_\varphi \stackrel{\rightarrow}{W}{}^{\!\!E} u_\varphi$.
Applying the Wronskian operator to (\ref{homogen1}) we get
	\begin{eqnarray}
	\Delta_{\varphi} = \frac{6 \pi^2 (1+\delta)m_{P}^{2}}
	{\lambda \varphi_{0}^{2}} \left[\,1
	+ O(m_P^2/|\xi|\varphi_{0}^{2})\,\right],\label{deltahomogen1}
	\end{eqnarray}
so that the final contribution of the inflaton mode equals
	\begin{eqnarray}
	\ln\frac{\Delta_{\varphi}^{1/2}}{|v_{\varphi}(t)|}=
	{\rm const}-2\ln \varphi_0
	+O(m_P^2/|\xi|\varphi_{0}^{2}).     \label{inflatonmode}
	\end{eqnarray}

\section{One-loop effective action on the DeSitter instanton}
\hspace{\parindent}
In this section we begin calculating the contribution of the
one-loop effective action (\ref{4.3.4}) to the distribution function
(\ref{4.3.2}). As it was discussed in the end of Sect.2,
this quantity should be calculated at the extremal
$Q=Q^{(0)}+\delta Q$ which differs from the exact DeSitter
background (with constant inflaton) $Q^{(0)}$ by the corrections of the
slow roll expansion:
	\begin{eqnarray}
	{\mbox{\boldmath $\Gamma$}}_{\rm 1-loop}=
        \left.\frac12\,{\rm Tr\,ln}\,
        \frac{\delta^2 I[\,\xi\,]}{\delta \xi\,\delta \xi}
        \,\right |_{\,Q^{(0)}+\delta Q}=
	{\mbox{\boldmath $\Gamma$}}_{\rm 1-loop}^{(0)}+
	\delta{\mbox{\boldmath $\Gamma$}}_{\rm 1-loop}.     \label{5.1}
	\end{eqnarray}
So here we obtain ${\mbox{\boldmath $\Gamma$}}_{\rm 1-loop}^{(0)}$
and in the Sect.6 develop the perturbation theory for
$\delta{\mbox{\boldmath $\Gamma$}}_{\rm 1-loop}$.

It is important that in contrast to the 2-dimensional minisuperspace
sector $(a,\varphi)$ that was only probed by the tree-level
approximation of the theory, the one-loop order involves the contribution
of {\it all} fields inhabiting the model. Without loss of generality
we shall assume that its low-energy (sub-Planckian) sector
is given by the inflaton-graviton action (\ref{4.2.3}) plus
arbitrary set of Higgs scalars $\chi(x)$, vector gauge bosons $A_\mu(x)$
and fermions $\psi(x)$. It can also include gravitino, but we shall mainly
focus at this sector of spin 0, 1/2, 1 and spin 2 fields. In the full
Lagrangian
	\begin{eqnarray}
	&&{\mbox{\boldmath $L$}}(g_{\mu\nu},\varphi,\chi,A_\mu,\psi)=
	{\mbox{\boldmath $L$}}(g_{\mu\nu},\varphi)
	+g^{1/2}\left(-\frac 12 \sum_\chi (\nabla\chi)^2\right.\nonumber\\
	&&\qquad\qquad\qquad\qquad\qquad\qquad
	\left.-\frac 14 \sum_A F_{\mu\nu}^2(A)
	-\sum_\psi \bar{\psi}\hat{\nabla}\psi\right)
	+{\mbox{\boldmath $L$}}_{\rm int}
	(\varphi,\chi,A_\mu,\psi)      \label{partmodel}
	\end{eqnarray}
we single out the interaction of Higgs, vector and
spinor fields with the inflaton field
${\mbox{\boldmath $L$}}_{\rm int}(\varphi,\chi,A_\mu,\psi)$. Its
nonderivative part has the form
	\begin{eqnarray}
	{\mbox{\boldmath $L$}}_{\rm int}
	=\sum_{\chi}\frac{\lambda_{\chi}}4
	\chi^2\varphi^2
	+\sum_{A}\frac12 g_{A}^2A_{\mu}^2\varphi^2+
        \sum_{\psi}f_{\psi}\varphi\bar\psi\psi
	+{\rm derivative\,\,coupling}             \label{interaction}
	\end{eqnarray}
with Higgs $\lambda_\chi$, vector gauge $g_A$ and Yukawa $f_\psi$
coupling constants. In the Lagrangian (\ref{partmodel}) the inflaton
field can be regarded as one of the components of one of the Higgs
multiplets $\chi$, which has a nonvanishing expectation value in the
cosmological quantum state. In its turn the choice of the interaction
Lagrangian here is dictated by the renormalizability of the matter field
sector of the theory (\ref{partmodel}) and by the requirement of local
gauge invariance with respect to arbitrary Yang-Mills group of vector
fields $A_\mu$. The terms of derivative coupling in (\ref{interaction})
should be chosen to guarantee the latter, but their form will not be
important for the conclusions of this paper. On the contrary, the
quantum gravitational effects will crucially depend on the nonderivative
part of the interaction Lagrangian.

A very important property of the functional (\ref{5.1}) is that it is
calculated {\it on shell}, that is on the solution of classical equations,
and therefore, as is well known from the theory of gauge fields
\cite{DeWitt,Bar-Vilk,Bar-Kam-Tun},
is independent of the choice of gauge conditions used for its construction
(or ADM reduction to physical variables in terms of which it reduces to
the functional determinant in the physical sector). This freedom can be used
to transform (\ref{5.1}) identically to the background covariant
gauges in which the one-loop action takes the form of the functional
determinant of the covariant operator
	\begin{eqnarray}
	\mbox{\boldmath$F$}=
	\frac{\delta^{2}{\mbox{\boldmath$I$}}_{\rm tot}
	[\,\mbox{\boldmath$g$}\,]}
	{\delta {\mbox{\boldmath$g$}}
	\,\,\delta {\mbox{\boldmath$g$}}}             \label{F-define}
	\end{eqnarray}
acting in the full space of gauge and ghost fields
	\begin{eqnarray}
	{\mbox{\boldmath$g$}}=(\varphi(x),\,\chi(x),\,
	\psi(x),\,A_\mu(x),\,
	\psi_\mu(x),\,g_{\mu\nu}(x),\,...,
	C(x),\,\bar{C}(x)).                 \label{5.2}
	\end{eqnarray}
Here $\mbox{\boldmath$I$}_{\rm tot}[\,\mbox{\boldmath$g$}\,]$ is
a total action defined on this space of fields including the covariant
gauge-breaking and ghost terms, $C$'s denote all possible gauge and
coordinate ghosts. Spatial components of fields in the nonghost sector
of (\ref{5.2}) form the canonical superspace of the theory (\ref{4.2.6}).
As compared to (\ref{4.2.6}) we only added $\chi(x)$ -- the set of all
scalar multiplets of the model other than the inflaton field $\varphi(x)$,
introduced above.

The possibility to convert the one-loop action to covariant form is very
important, because in this form it admits covariant regularization
and renormalization of inevitable ultraviolet divergences and allows one
to obtain correct scaling behaviour, quantum anomalies, etc. The full
set of gauges for internal gauge symmetries and gravitational diffeomorphisms
can be chosen in such a way that the operator (\ref{F-define}) becomes
minimal -- diagonal in second-order derivatives forming a covariant
D'Alambertian $\Box$. Moreover, on the exact DeSitter background with
constant inflaton field this operator can as a whole be reduced to the
block-diagonal form
	\begin{eqnarray}
	\mbox{\boldmath$F$}={\rm diag}\,(-\Box_s+X_s)  \label{block}
	\end{eqnarray}
with blocks $-\Box_s+X_s$ belonging to $O(5)$ irreducible representations
of spin $s=0,\,1/2,\,1,\,3/2,\,2,\,...$ on the 4-dimensional sphere
of the Euclidean DeSitter space. For constant potential terms $X_s$ of these
operators (which is the case of $Q^{(0)}$) their spectra are
well known \cite{Christensen-Duff,Allen,Frad-Tseyt} and can be used for
calculating their functional determinants under a proper covariant
regularization. We shall use a $\zeta$-functional regularization
\cite{zetafunction} in which a contribution of every block of (\ref{block})
equals
	\begin{eqnarray}
	-\frac 12 {\rm Tr}\,\ln (-\Box_s+X_s)=
	\frac{1}{2} \zeta'_s(0) +\frac{1}{2} \zeta_s(0)
	\ln \frac{\mu^2}{H^2},       \label{one-loopaction}
	\end{eqnarray}
with the generalized $\zeta$-functions built of dimensionless
eigenvalues of the rescaled operator
	\begin{eqnarray}
	\zeta_s(p)=\sum_\lambda \lambda^{-p},\,\,\,\,
	H^{-2}(-\Box_s+X_s)\phi_\lambda(x)
	=\lambda\phi_\lambda(x)      \label{zetas}
	\end{eqnarray}
(the scale of the DeSitter instanton is its inverse radius $H$, so that
the differential operator here is effectively defined on a sphere of unit
radius with a correspondingly rescaled potential term). As a result
the full one-loop action takes the form
	\begin{eqnarray}
	&&-{\mbox{\boldmath $\Gamma$}}_{\rm 1-loop}^{(0)}
	= \frac{1}{2}\zeta'(0)
	+\frac{1}{2} \zeta(0) \ln
	\frac{\mu^2}{H^2},                     \label{gammazet}\\
	&&\zeta'(0)=\sum_s w_s \zeta'_s(0),\,\,\,
	\zeta(0)=\sum_s w_s \zeta_s(0).      \label{totzet}
	\end{eqnarray}
Here the weights $w_s$ -- positive and negative integers -- reflect
the statistics of the field and also the details of
transition from the original local fields (\ref{5.2}) to the decomposition in
irreducible $O(5)$ representations of (\ref{block})
\footnote
{This transition involves certain Jacobians which have the form of positive
and negative powers of operators in (\ref{block})
\cite{Christensen-Duff,Allen,Frad-Tseyt} leading to sign factors of the above
type. The example of this procedure is presented below for gauge vector field.
}.

In the equations above $\mu^2$ is a mass parameter reflecting
the renormalization ambiguity resulting from the subtraction of
logarithmic divergences proportional to $\zeta(0)$. This quantity plays
a very important role because it determines the leading high-energy
behaviour of the one-loop action and correspondingly the anomalous
scaling behaviour of the distribution function $\rho\sim H^{-\zeta(0)}$.
As it was observed in
\cite{BKam:norm,Bar-Kam-Scale,Bar-Kam-Tun,QGrav,AltBarv} it can
produce a principal quantum cosmological mechanism -- to make
the distribution function of quasi-DeSitter models normalizable in
over-Planckian domain and generate the inflationary probability peak.
Therefore we begin with the calculation of this quantity.

\subsection{The anomalous scaling $\zeta(0)$}
\hspace{\parindent}
Apart from the method of $O(5)$ irreducible representations and Eqs.
(\ref{zetas}) and (\ref{totzet}) the total anomalous scaling $\zeta(0)$
can be obtained by a more universal Schwinger-DeWitt technique
\cite{DeWitt,Bar-Vilk} as an integral over the spacetime
	\begin{eqnarray}
	\zeta(0)=\frac 1{16\pi^2}\int d^4x\,g^{1/2}\, a_2(x) \label{5.2.1}
	\end{eqnarray}
of the second Schwinger-DeWitt coefficient $a_2(x)$ in the proper time heat
kernel expansion for the operator (\ref{F-define}). This technique easily
allows one to derive the mechanism of suppression of the over-Planckian
energy scales due to a big
value of the nonminimal coupling constant $|\xi|\gg 1$. Indeed, for
large masses of particles the dominant contribution to $a_2(x)$
	\begin{eqnarray}
	a_2(x)=\frac 12 \left(\sum_\chi m_\chi^4+4\sum_A m_A^4-
	4\sum_\psi m_\psi^4\right)+...            \label{5.2.2}
	\end{eqnarray}
is quartic in their masses with the sign factor depending
on statistics (and weight factors given by the number of the corresponding
tensor field components). In the model (\ref{partmodel})-(\ref{interaction})
on the background with a big constant field $\varphi_0$
Higgs scalars, vector gauge bosons and fermions acquire by the analogue of
the Higgs mechanism the effective masses induced by the interaction
Lagrangian
	\begin{eqnarray}
	m_{\chi}^{2} = \frac{\lambda_\chi \varphi_0^2}{2},\,\,
	m_A^2=g_A^2\varphi_0^2,\,\,
	m_\psi^2=f_\psi^2\varphi_0^2.     \label{effectivemass}
	\end{eqnarray}
Beeing integrated over the 4-volume of the instanton $8\pi^2/3H^4$ these
masses generate in view of the expression (\ref{Hubble}) for $H(\varphi_0)$
the following dominant contribution to $\zeta(0)$
	\begin{eqnarray}
	\zeta(0)=Z\,[\,1+O(m_p^2/|\xi|\varphi_0^2)\,],\,\,\,
	Z=6\,\frac{\xi^{2}}{\lambda}
	{\mbox{\boldmath $A$}},       \label{5.2.3}
	\end{eqnarray}
where ${\mbox{\boldmath $A$}}$ is a following fundamental combination of
matter fields coupling constants \cite{Bar-Kam-Scale}
	\begin{eqnarray}
	{\mbox{\boldmath $A$}} = \frac{1}{2\lambda}
	\left(\sum_{\chi} \lambda_{\chi}^{2}
	+ 16 \sum_{A} g_{A}^{4} - 16
	\sum_{\psi} f_{\psi}^{4}\right).   \label{fieldcontent}
	\end{eqnarray}
Thus for large $|\xi|\gg 1$ and positive ${\mbox{\boldmath $A$}}\sim O(1)$
we have $Z\gg 1$ which is a corner stone of the quantum gravitational
mechanism that suppresses the over-Planckian energy scales,
$\rho\sim H^{-Z}\rightarrow 0,\,H\rightarrow\infty$ and, thus, serves as
a justification of the semiclassical expansion.

It is important that this mechanism is entirely generated in the
renormalizable matter-field sector of the theory consisting of the
multiplets of the standard model, because the graviton-inflaton sector
(\ref{4.2.3}) yields the contribution independent of $\xi$
\cite{Gryz-Kam-Karm,BKK} as well as the spin-3/2 gravitino field
\cite{Poletti} (considered to be noninteracting with inflaton)
	\begin{eqnarray}
	\zeta(0)_{\rm graviton +
	inflaton}=-\frac{171}{10},\,\,
	\zeta(0)_{\rm gravitino}=\frac{589}{180}.   \label{grav+infl2}
	\end{eqnarray}
This property suggests that this mechanism is robust
against high energy modifications of the fundamental theory designed
to solve the problems of nonrenormalizability in perturbative quantum
gravity
\footnote{Another advantage of this mechanism related to big $|\xi|$ is
that it allows one to disregard the known and thus far unresolved problem
of discrepancies between the renormalization in covariant and unitary
gauges observed on topologically nontrivial curved spaces for fields
of spins $s\geq 1$ \cite{discrepancies1,discrepancies2,discrepancies3}.
These discrepancies are independent of the nonminimal coupling constant
and, thus, negligible for $|\xi|\gg 1$.
}.

Our purpose now is to go beyond the lowest order approximation (\ref{5.2.3})
and find subleading corrections. For this we would need the subleading term
in the expression for the Hubble constant (\ref{Hubble}) and also use
exact expressions for $\zeta(0)$ of massive scalar, vector and spinor
fields on the Euclidean DeSitter background. For a scalar field with the
mass $m_\chi$ and the constant $\xi_\chi$ of nonminimal interaction
(which is different from $\xi\equiv\xi_\varphi$) this expression reads
\cite{Christensen-Duff,Allen,Frad-Tseyt}
	\begin{eqnarray}
	\zeta_\chi(0)=\frac{29}{90}-4\,\xi_\chi+12\,\xi_\chi^{2}
	-\frac{1}{3}\,\frac{m_\chi^2}{H^2}
	+\frac{1}{12}\,\frac{m_\chi^4}{H^4}.    \label{Z-scal}
	\end{eqnarray}
Using (\ref{Hubble}) and (\ref{effectivemass}) in this equation we have
	\begin{eqnarray}
	\zeta_\chi(0)= \frac{3\lambda_{\chi}^2}{\lambda^2}\,
	\xi^2  + \frac{3\lambda_\chi^2(1 +2\delta)}{4\pi\lambda^2}
	\frac{|\xi|\,m_P^2}{\varphi_0^2}
	-\frac{2\lambda_{\chi}}
	{\lambda}\,|\xi| + O(m_P^2/\varphi_0^2).    \label{Z-scal1}
	\end{eqnarray}
The structure of this expression
	\begin{eqnarray}
	O(|\xi|^2)+O(|\xi|m_P^2/\varphi_0^2)
	+O(|\xi|)+O(m_P^2/\varphi_0^2)       \label{5.2.4}
	\end{eqnarray}
demonstrates the nature of the perturbation theory that we shall use in
what follows. It has two smallness parameters: $m_P^2/|\xi|\varphi_0^2\ll 1$
-- the parameter of the slow roll expansion and $1/|\xi|\ll 1$ -- the
parameter of this particular model with nonminimal inflaton. Below  we
shall see that $|\xi|\simeq 2\times 10^4$ and the probability maximum
in the distribution in question will be for $m_P/\varphi_0\simeq 0.03$.
This means that for the most important range of values of the inflaton field
	\begin{eqnarray}
	\frac {m_P^2}{|\xi|\varphi_0^2}\gg\frac 1{|\xi|}
	\end{eqnarray}
and in the subleading approximation the last two terms in (\ref{Z-scal1})
and (\ref{5.2.4}) can be discarded. In what follows we shall invariably
follow this rule. Thus we have
	\begin{eqnarray}
	\zeta_\chi(0)=3\xi^{2}\,
	\frac{\lambda_{\chi}^{2}}{\lambda^{2}}
	\left[\,1 + \frac{1 +
	2\delta}{4\pi}\frac{m_P^2}
	{|\xi|\varphi_{0}^{2}} +
	O(1/|\xi|)\, \right].                         \label{Z-scal2}
	\end{eqnarray}

Similar calculations for vector and Dirac spinor fields with masses from
(\ref{effectivemass}) give the result (taking into account their
statistics)
	\begin{eqnarray}
	&&\zeta_A(0)= 48\,\xi^{2}\, \frac{g_{A}^{4}}{\lambda^{2}}
	\left[\,1 + \frac{1 +
	2\delta}{4\pi}\frac{m_P^2}
	{|\xi|\varphi_{0}^{2}} +
	O(1/|\xi|)\, \right],               \label{Z-vect}\\
	&&\zeta_\psi(0) = -48\,\xi^{2}\,\frac{f_\psi^2}{\lambda^2}
	\left[\,1 + \frac{1 +
	2\delta}{4\pi}\frac{m_P^2}
	{|\xi|\varphi_{0}^{2}} +
	O(1/|\xi|)\, \right].                 \label{Z-Dirac}
	\end{eqnarray}
Obviously, the contribution of Majorana or Weyl spinor fields is one half
of the expression (\ref{Z-Dirac}).

Thus, the total anomalous scaling of the theory on the
exact DeSitter background reads
	\begin{eqnarray}
	\zeta(0)=6\,\frac{\xi^{2}}{\lambda} {\mbox{\boldmath $A$}}
	\left[\,1 + \frac{1 +
	2\delta}{4\pi}\frac{m_P^2}
	{|\xi|\varphi_{0}^{2}} +
	O(1/|\xi|)\, \right].                 \label{Z-total}
	\end{eqnarray}

It should be emphasized again that the graviton-inflaton sector does
not contribute to this expression. We have seen that dependence on
$\xi$ in $\zeta_\chi(0)$ above arises due to the terms $m_\chi^{4}/H^{4}$ and
$m_\chi^{2}/H^{2}$. Indeed, in these terms $H$ depends on $\xi$. However,
while the effective masses $m_\chi,\,m_A,\,m_\psi$ of non-inflaton scalar,
vector and spinor fields do not depend on $\xi$, the effective masses
of the graviton and inflaton are $\xi$-dependent (see eq.(\ref{effectmass1})
for the effective inflaton mass) and this dependence
cancels the dependence of $H$ on $\xi$. One can show also that the subleading
terms in $\zeta(0)_{\rm gravity + inflaton}$ will be at most $O(|\xi|)$ in
the terminology of (\ref{5.2.4}) and thus will be discarded in what follows
From calculations below it will be clear that the same is true
for $\zeta'(0)$ too.  Thus in the subleading approximation
the one-loop part of the probability distribution will be contributed by
the non-inflaton scalar, vector and spinor fields.

\subsection{Calculating $\zeta'(0)$}
\hspace{\parindent}
The calculation of $\zeta'(0)$ requires the knowledge of the finite part
of the effective action unrelated to its ultraviolet divergences associated
with $\zeta(0)$. We shall calculate it by the technique of generalized
$\zeta$-functions developed in \cite{Allen,Frad-Tseyt} for fields in the
basis of irreducible $O(5)$ representations of different spins $s$
(\ref{block}). In \cite{Frad-Tseyt} these calculations for $\zeta$-functions
(\ref{zetas}) were put in a unified framework in the form of a special
function defined for $\Re{\rm e}\,p>2$ and analytically continued to
$p=0$
	\begin{eqnarray}
	&&\zeta_{s}(p) = \frac{1}{3} (2s+1) F
	\left(p,2s+1,(s+1/2)^2,b_s\right),      \label{zetaviaF}\\
	&&F(p,k,a,b) \equiv \sum_{\nu=\frac{1}{2} k +1}^{\infty}
	\frac{\nu(\nu^{2} - a)}
	{(\nu^{2}-b)^{p}},\,\,\,\,\Re{\rm e}\,  p > 2.   \label{defineF}
	\end{eqnarray}
Here the parameters $b_{s}$ are related to potential terms of operators
in (\ref{block}) according to \cite{Frad-Tseyt}:
	\begin{eqnarray}
	b_0 = \frac{9}{4} - \frac{X_0}{H^2},\,
	b_1 = \frac{13}{4}-\frac{X_1}{H^{2}},\,
	b_{1/2} = -\frac{X_{1/2}}{H^{2}}.     \label{define-spin-b}
	\end{eqnarray}
Particular values of the function (\ref{defineF}) equal \cite{Frad-Tseyt}
	\begin{eqnarray}
	&&F(0,k,a,b) = \frac{1}{4} b(b-2a)
	+\frac{1}{24}a(3k^{2}+6k+2)-
	\frac{1}{64}k^{2}(k+2)^{2}
	+\frac{1}{120},                      \label{F0}\\
	&&F(1,k,a,b) = \frac{1}{2} b -\frac{1}{12}
	-\frac{1}{8} k(k+2)
	-\frac{1}{2}(b-a)\Psi\left(\frac{k}{2}+1
	\pm \sqrt{b}\right),                      \label{F1}\\
	&&\Psi(x \pm y) \equiv \psi(x+y) + \psi(x-y),
	\end{eqnarray}
where $\psi(x)$ is a logarithmic derivative of Euler's $\Gamma$ function.
Using the corollary of (\ref{defineF})
	\begin{eqnarray}
	\frac{d}{db} F'(0,k,a,b) = F(1,k,a,b),
	\,F'=\frac{dF}{dp},                     \label{defineF'}
	\end{eqnarray}
we can find $F'(0)$ and hence $\zeta'(0)$ by integrating (\ref{defineF'})
	\begin{eqnarray}
	&&F'(0,k,a,b) = \frac{1}{4} b^{2} -\frac{1}{12}b
	-\frac{1}{8}b k(k+2)                            \nonumber \\
	&&\qquad\qquad\qquad\qquad-\frac{1}{2}\int_{0}^{b} dz(z-a)
	\Psi\left(\frac{k}{2}
	+1 \pm\sqrt{z}\right)+ C,               \label{defineF'1}
	\end{eqnarray}
where the constant $C$ is given by derivatives of Hurwitz functions
\[C = 2\zeta'_{H}(-3,1+k/2)-2a\zeta'_{H}(-1,1+k/2),\,\,
\zeta_{H}(x,y) = \sum_{n=0}^{\infty} \frac{1}{(n+y)^{x}}.\]

The expression (\ref{defineF'1}) is rather complicated and generally can be
obtained only numerically. However, we need only its dependence on
$\varphi_0$ contained in $b_s$ and in the approximation retaining only the
first two terms of the generic expansion (\ref{5.2.4}). Since the potential
terms of (\ref{block}) are basically given by effective masses induced
according to (\ref{effectivemass}), $X_s\sim m_s^2=O(\varphi_0^2)$, the
corresponding parameters $b_s$ (\ref{define-spin-b}) have the structure
	\begin{eqnarray}
	b_s = b_s^{(0)} + \frac{b_s^{(1)}}{\varphi_{0}^{2}}+O(1/|\xi|),\,\,
	b_s^{(0)}\simeq -\frac{X_s}{H^2}=O(|\xi|).    \label{b-expansion}
	\end{eqnarray}
Using (\ref{F0})-(\ref{F1}) and expanding (\ref{zetaviaF}) and
(\ref{defineF'1}) in $b_s^{(1)}/\varphi_0^2$ one obtains
	\begin{eqnarray}
	&&\zeta_s(0)=\left.\zeta_s(0)\right|_{b_s^{(0)}}
	+\frac{2s+1}6 \frac{b_s^{(0)}b_s^{(1)}}{\varphi_0^2}+
	O(m_P^2/\varphi_0^2),                              \label{5.2.8}\\
	&&\zeta_s'(0) =\left.\zeta_s'(0)\right|_{b_s^{(0)}}+
	\frac{2s+1}{3} \left\{\,\frac{b_s^{(0)}}{2}\right.
	\left.-\frac12 \left[b_s^{(0)}
	-(s+1/2)^2\right]
	\Psi\left(s+\frac 32 \pm i\sqrt{-b_s^{(0)}}\right)\right\}\nonumber\\
	&&\qquad\qquad\qquad\qquad\qquad\qquad\qquad\qquad
	\qquad\qquad\qquad\qquad
	+O\left(\frac{m_P^2}{\varphi_0^2}\right).   \label{zeta'-on-scal}
	\end{eqnarray}
For the calculation of the full effective action (\ref{one-loopaction})
we would also need $\ln(\mu^2/H^2)$. It is worth transforming this quantity
separately for every irreducible $s$-component
	\begin{eqnarray}
	\ln \frac{\mu^{2}}{H^{2}} =
	\ln\frac{\mu^2}{\varphi_0^2}+
	\ln\frac{\varphi_0^2}{X_s}+\ln(-b_s^{(0)})+
	\frac{b_s^{(1)}}{b_s^{(0)}\varphi_0^2}
	+O\left(\frac 1{|\xi|}\right).            \label{log-Hubble}
	\end{eqnarray}
Then, disregarding the $\varphi_0$-independent part and using in
$\Psi\left(s+3/2 \pm i(-b_s^{(0)})^{1/2}\right)$
the asymptotic expansion for $\psi(z)$ at large $z$
\cite{Abra-Sti}, $\psi(z) \sim \ln z -1/2z +\ldots$, we see that the
logarithmic in $b_s^{(0)}$ term of (\ref{log-Hubble}) gets cancelled
by the logarithmic term of $\Psi\left(s+3/2 \pm i(-b_s^{(0)})^{1/2}\right)$
and the final answer for a partial one-loop action reads
	\begin{eqnarray}
	&&\zeta_s'(0) +\zeta_s(0) \ln \frac{\mu^2}{H^{2}} =
	{\rm const} +\zeta_s(0)\ln \frac{\mu^2}{\varphi_0^2}\nonumber\\
	&&\qquad\qquad\qquad\qquad\qquad\qquad
	+ \frac{2s+1}{3}\frac{b_s^{(1)}b_s^{(0)}}{\varphi_{0}^{2}}
	\left(\frac{3}{4} +\frac{1}{2}\ln \frac{\varphi_0^2}{X_s}
	\right)
	+O\left(\frac{m_P^2}{\varphi_0^2}\right).  \label{one-loop-contr1}
	\end{eqnarray}

Now we can go over to the calculation of the total one-loop action for the
model (\ref{partmodel}). For a Higgs scalar field $\chi$ the contribution
reduces to the above equations (\ref{5.2.8}) and (\ref{one-loop-contr1})
with $s=0$ and $X_0=m_\chi^2$ (see eq.(\ref{effectivemass}) for
the effective mass of the Higgs field), so that in view of the expansion
for $H$ (\ref{Hubble})
	\begin{eqnarray}
	b_\chi^{(0)} = -6\,|\xi|\frac{\lambda_{\chi}}{\lambda},\,
	b_{1 {\rm scal}} = -\frac{3\lambda_{\chi}
	(1+2\delta)}{4\pi\lambda}m_P^2.     \label{b-scal}
	\end{eqnarray}
For a gauge vector field the situation is more complicated because its
one-loop action in the most convenient gauge $\nabla^{\mu} A_{\mu}=0$
(leading to minimal operator with diagonal derivatives) equals
\cite{Frad-Tseyt}
	\begin{eqnarray}
	&&\frac 12\,{\rm Tr}\,\ln (-\Box \delta^\mu_\nu+R^\mu_\nu+m_A^2)
	-{\rm Tr}\,\ln (-\Box_0)=
	\frac 12\,{\rm Tr}\,\ln (-\Box_1+m_A^2+3H^2)\nonumber\\
	&&\qquad\qquad\qquad\qquad\qquad
	+\frac 12\,{\rm Tr}\,\ln (-\Box_0+m_A^2)
	-{\rm Tr}\,\ln (-\Box_0),     \label{vectorstructure}
	\end{eqnarray}
where the subtracted term is a contribution of ghosts and the other
terms represent the decomposition of the vector functional determinants
into $O(5)$ irreducible components (see footnote after eq.(\ref{totzet})).
Similar procedure holds for a spinor field operator which technically must be
squared to reduce calculations to that of the $\Box_{1/2}+...$
\cite{Frad-Tseyt}. Finally we have for vector and spinor fields the
coefficients
	\begin{eqnarray}
	&&b_\psi^{(0)} = -12\,|\xi| \frac{f_{\psi}^{2}}{\lambda},\,
	b_\psi^{(1)} = -\frac{3f_{\psi}^{2} (1+2\delta)}
	{2\pi\lambda}m_P^2,                   \label{b-Dirac}\\
	&&b_A^{(0)} = - 12\,|\xi| \frac{g_{A}^{2}}{\lambda},\,
	b_A^{(1)} =-\frac{3g_{A}^{2}(1+2\delta)}{2\pi\lambda}
	m_P^2                                       \label{b-vect}
	\end{eqnarray}
and the corresponding $X_A,\,X_\psi$ different from the squared masses
(\ref{effectivemass}) by terms that go beyond our approximation. Their use
in (\ref{5.2.8}) and (\ref{one-loop-contr1}) allows one to reproduce
the expression (\ref{Z-total}) for the total $\zeta(0)$ obtained above by
the Schwinger-DeWitt method and get the final algorithm for the total
one-loop effective action on the DeSitter background
	\begin{eqnarray}
	&&-{\mbox{\boldmath $\Gamma$}}_{\rm 1-loop}^{(0)}
	= {\rm const}-3\,\frac{\xi^{2}}{\lambda} {\mbox{\boldmath $A$}}
	\left[\,1 + \frac{1 +
	2\delta}{4\pi}\frac{m_P^2}
	{|\xi|\varphi_{0}^{2}}\, \right]
	\ln\frac{\varphi_0^2}{\mu^2} \nonumber \\
	&&\qquad\qquad\qquad\qquad\qquad
	+\frac{3|\xi|(1+2\delta)m_{P}^{2}}{4\pi\lambda\varphi_{0}^{2}}
	\left(\frac{3}{2}{\mbox{\boldmath $A$}}
	+ {\mbox{\boldmath $B$}}\right)
	+O\left(\frac{m_P^2}{\varphi_0^2}\right), \label{one-loop-contr3}
	\end{eqnarray}
where the coefficient ${\mbox{\boldmath $A$}}$ is defined
by  Eq. (\ref{fieldcontent}) and ${\mbox{\boldmath
$B$}}$ is a following new combination of coupling constants
	\begin{eqnarray}
	{\mbox{\boldmath $B$}} =
	-\frac{1}{2\lambda} \left(\sum_{\chi} \lambda_{\chi}^{2} \ln
	\frac{\lambda_{\chi}}{2}
	+ 16\sum_{A} g_{A}^{4} \ln g_{A}^{2}
	-16\sum_{\psi} f_{\psi}^{4}
	\ln f_{\psi}^{2}\right).      \label{fieldcontent1}
	\end{eqnarray}

\section{Perturbation theory for the one-loop effective action}
\hspace{\parindent}
Due to the presence of slow roll corrections $\delta Q$ the effective
action (\ref{5.1}) acquires the contribution which in the first order
of the perturbation theory equals
	\begin{eqnarray}
	&&\delta{\mbox{\boldmath $\Gamma$}}_{\rm 1-loop}=
	\left.\frac{1}{2}\, {\rm Tr}\,
	[\,\delta \mbox{\boldmath$F G$}\,]\,
	\right|_{Q^{(0)}},         \label{vareffectaction}\\
	&&\mbox{\boldmath$F G$}(x,x') = \delta(x,x')
	\end{eqnarray}
where $\mbox{\boldmath$G$}=\mbox{\boldmath$G$}(x,x')$
is the Green's function of the operator $\mbox{\boldmath$F$}$
(eq.(\ref{F-define})) on a four-sphere and $\delta \mbox{\boldmath$F$}$
is the variation of this operator induced by
$\delta Q = (\delta a,\delta\varphi)$.

The expression (\ref{vareffectaction}) is incomplete
unless one fixes uniquely the Green's function and specifies the functional
composition law $\delta\mbox{\boldmath$F G$}$ in the functional trace. One
should remember that the kernel $\mbox{\boldmath$G$}(x,x')$ is not
a smooth function of its arguments and its
irregularity enhances when it is acted upon by two derivatives contained
in $\delta\mbox{\boldmath$F$}$. Therefore one has to prescribe the way
these derivatives act on both arguments of $\mbox{\boldmath$G$}(x,x')$ and
how the coincidence limit of the resulting singular kernel is taken
in the functional trace
	\begin{eqnarray}
	{\rm Tr}\,[\,\delta\mbox{\boldmath$F G$}\,]=
	\left.\int d^4x\,{\rm tr}\,
	\delta \mbox{\boldmath$F$}(\nabla',\nabla)
	\mbox{\boldmath$G$}(x,x')\,\right|_{x'=x}    \label{6.1}
	\end{eqnarray}
(${\rm tr}$ denotes the matrix trace operation over tensor indices).
The specification of trace operation follows from the procedure of
calculating the Gaussian path integral over quantum disturbances
which gives rise to the one-loop functional determinants.
As was shown in \cite{Feynman} (see also \cite{DeWitt1} and
\cite{reduct}) the functional determinant of the differential
operator generated by the Gaussian path integral is determined by the
variational equation $\delta \ln {\rm Det}\, \mbox{\boldmath$F$} \equiv
\delta {\rm Tr}\,\ln \mbox{\boldmath$F$}
={\rm Tr}\,\delta\mbox{\boldmath$F\,G$}$,
where the Green's function $\mbox{\boldmath$G$}(x,x')$ satisfies the same
boundary conditions as the integration variables in the Gaussian integral
and the functional composition law $\delta\mbox{\boldmath$F\,G$}$ implies a
symmetric action of spacetime derivatives on both arguments of
$\mbox{\boldmath$G$}(x,x')$. In what follows we shall describe in detail
the calculations for the case of the scalar field and then give the
result for other higher spin contributions.

For the field $\chi(x)$ with the wave operator
	\begin{eqnarray}
	\mbox{\boldmath$F$} = -g^{\mu\nu} \nabla_{\mu}\nabla_{\nu}
	+ m_\chi^{2}(\varphi_0)    \label{Klein-Gordon}
	\end{eqnarray}
the symmetrized variation of this operator (\ref{Klein-Gordon}) looks like
	\begin{eqnarray}
	\delta \mbox{\boldmath$F$}(\nabla',\nabla) =
	\nabla_{\mu'} \delta g^{\mu\nu}(x)\nabla_{\nu}
	+ \delta m_\chi^{2}(\varphi),  \label{Klein-Gordon-var}
	\end{eqnarray}
where $\nabla_{\mu'}$ and $\nabla_{\nu}$ are acting on different
arguments of $\mbox{\boldmath$G$}(x,x')$ in (\ref{6.1})
\footnote
{Note that $\mbox{\boldmath$F$}$ enters the action with the metric dependent
factor $g^{1/2}\mbox{\boldmath$F$}$, however $g^{1/2}$ is not varied here,
because the contribution of this overall factor is cancelled by the
contribution of the local measure \cite{FV:Bern}. For this reason, in
particular, the effective action is given by the functional determinants of
$\mbox{\boldmath$F$}$ instead of those of $g^{1/2}\mbox{\boldmath$F$}$.
The difference between the corresponding results as well as the contribution
of local measure are formally proportional to unregulated $\delta(0)$
which vanishes in dimensional regularization, but given by $\zeta(0)$ in
the zeta-functional one. Therefore in the regularization we use here
these terms require a careful bookkeeping.
}.

The Green's function for a scalar field on a four-sphere is well known
\cite{Green's}, however we have to regulate a divergent expression
(\ref{6.1}) and will do it by replacing the Green's function with
its operator power
$\mbox{\boldmath$G$}^{1+p}(x,x')=\mbox{\boldmath$F$}^{-1-p}\delta(x,x'),\,
p\rightarrow 0$,
	\begin{eqnarray}
	\mbox{\boldmath$G$}^{1+p}(x,x')= \frac{1}{\Gamma(p+1)}
	\int _{0}^{\infty} ds s^{p}
	\exp(-s\mbox{\boldmath$F$}) \delta(x,x').    \label{Green's1}
	\end{eqnarray}
The corresponding heat kernel can be constructed by noting that in view of
DeSitter invariance both the Green's function and its heat kernel are
functions of the world function $\sigma(x,x')$ -- one half of the square
of geodetic distance between the points $x$ and $x'$ which can also be
expressed in terms of the angle $\theta=\theta(x,x')$ between these points
on a sphere of radius $R=1/H$, $\sigma(x,x')=R^2\theta^2/2$. Similarly,
the dedensitized delta-function above can be constructed in terms of this
angular variable $y = \cos \theta$
	\begin{eqnarray}
	\frac{\delta(x,x')}{g^{1/2}(x')}=
	\frac{1}{4\pi^{2}R^{4}} \frac{d}{dy}
	\delta (y-1).                             \label{deltafunction}
	\end{eqnarray}
The scalar D'Alambertian acting on this function of $\theta$ yields the
operator of the Legendre equation after the commutation with the derivative
$d/dy$ above
	\begin{eqnarray}
	\Box\,\frac{d}{dy}=
	\frac{d}{dy}\left[(1-y^{2})\frac{d^{2}}{dy^{2}}
	-2y\frac{d}{dy}+2\right],
	\end{eqnarray}
which suggests to expand the delta function in the series of
Legendre polynomials $P_{n}(y)$ -- eigenfunctions of the Legendre
operator with eigenvalues $-n(n+1)$
	\begin{eqnarray}
	\delta (y-1) = \sum_{n=0}^{\infty}
	\left(n+\frac{1}{2}\right)P_{n}(y),   \label{deltafunction1}
	\end{eqnarray}
whence it follows that
	\begin{eqnarray}
	\exp (-s\mbox{\boldmath$F$})\delta(x,x') =
	\frac{g^{1/2}(x')}{4 \pi^{2} R^{4}} \sum_{n=1}^{\infty}
	\left(n+\frac{1}{2}\right) \,e^{-s\left[
	(n+1/2)^{2}-b_0\right]/R^2}
	dP_{n}(y)/dy,     \label{heatkernel1}
	\end{eqnarray}
where $b_0=9/4-m_\chi^2R^2$ coincides with the expression given by
(\ref{define-spin-b}) for a spin-$0$ case. Substituting it into (\ref{Green's1}),
integrating over $s$ and expressing the Legendre polynomials
in terms of the hypergeometric function we finally obtain
	\begin{eqnarray}
	&&\mbox{\boldmath$G$}^{1+p}(x,x')=
	\frac{g^{1/2}(x')}{16 \pi^{2} R^{2-2p}}
	\sum_{n=1}^{\infty} \frac{(-1)^{n+1}n(n+1)(2n+1)}
	{\left[(n+1/2)^{2}-b_0\right]^{1+p}}  \nonumber\\
	&&\qquad\qquad\qquad\qquad\qquad
	\times F\left(1-n,n+2;2;\frac 12+\frac 12
	\cos\sqrt{2\sigma(x,x')/R^2}\right).         \label{Green's2}
	\end{eqnarray}

This expression allows one to obtain the coincidence limits of the Green's
function and its derivatives arising in (\ref{6.1}). In view of the
well known coincidence limits
$\sigma(x,x)=0,\,\nabla_\mu\sigma(x,x')|_{x'=x}=0,\,
\nabla_\mu\nabla_{\nu'}\sigma(x,x')|_{x'=x}=-g_{\mu\nu'}$ we have
	\begin{eqnarray}
	&&\mbox{\boldmath$G$}^{1+p}(x,x) =
	\frac{g^{1/2}(x)}{16 \pi^{2} R^{2-2p}}
	\sum_{n=1}^{\infty} \frac{n(n+1)(2n+1)}
	{\left[(n+1/2)^{2}-b_0\right]^{1+p}}, \label{coincide}\\
	&&\left.\nabla_{\mu}\nabla_{\nu'}
	\mbox{\boldmath$G$}^{1+p}(x,x')\right|_{x=x'} =
	\frac{g_{\mu\nu}g^{1/2}(x)}{64 \pi^{2} R^{4-2p}}
	\sum_{n=1}^{\infty} \frac{n(n+1)(2n+1)(1-n)(n+2)}
	{\left[(n+1/2)^{2}-b_0\right]^{1+p}}. \label{coincide1}
	\end{eqnarray}
Infinite series here are calculable by the technique of
\cite{Christensen-Duff,Allen,Frad-Tseyt}. They have the
pole structure in $p\rightarrow 0$, $A/p+B+O(p)$, and lead
to $\delta{\mbox{\boldmath $\Gamma$}}_{\rm 1-loop}$ in the form
of the finite part of the variation (\ref{vareffectaction})
\footnote
{Note that $\zeta$-functional regularization is formally free from
divergences -- pole terms in $p$ \cite{zetafunction}. These terms are
artifacts of the variational equation (\ref{vareffectaction}) which
differs from the variation of finite $\zeta'(0)$
exactly by the pole term in $p$. Indeed $\delta\zeta'(p)=
-(1+p\,d/dp){\rm Tr}\,\delta\mbox{\boldmath$F$}\,\mbox{\boldmath$G$}^{1+p}$
and for ${\rm Tr}\,\delta\mbox{\boldmath$F$}\,\mbox{\boldmath$G$}^{1+p}=
A/p+O(1),\,p\rightarrow 0$, equals the finite part of
(\ref{vareffectaction}) $\delta\zeta'(0)=
\left[-{\rm Tr}\,\delta\mbox{\boldmath$F$}\,\mbox{\boldmath$G$}^{1+p}
+a/p\right]_{p=0}$.
}:
	\begin{eqnarray}
	\delta{\mbox{\boldmath $\Gamma$}}_{\rm 1-loop} =
	\frac{1}{2} (I_{1} A_{1} + I_{2} A_{2}) \ln \frac{\mu^2}{H^2}
	+ \frac{1}{2} (I_{1} B_{1} + I_{2} B_{2}),    \label{deltaW}
	\end{eqnarray}
where $I_{1}$ and $I_{2}$ are given by the integrals
	\begin{eqnarray}
	&&I_{1} = - \frac{1}{16 \pi^{2} R^{2}}
	\int_{S^{4}} d^{4} x g^{1/2}\delta m^{2},   \label{defineI1}\\
	&&I_{2} = \frac{1}{64 \pi^{2} R^{4}}
	\int_{S^{4}} d^{4} x g^{1/2}
	\delta g^{\mu\nu}g_{\mu\nu},          \label{defineI2}
	\end{eqnarray}
and $A_i,\,B_i,\,i=1,2$, are the following pole and finite parts of the
above two sums
	\begin{eqnarray}
	&&A_{1}=b_0-\frac14,\,\,B_{1}=\frac1{12}- \left(b_0-\frac14\right)
	\Psi\left(\frac12 \pm\sqrt{b_0}\right)+b_0, \nonumber\\
	&&A_{2}=\frac94 A_{1}-b_0\left(b_0-\frac14\right),\nonumber\\
	&&B_{2}=\frac94 B_{1}+\!\frac7{480}
	-\!\frac1{12}\left(b_0-\!\frac14\right)+\!
	b_0\left(b_0-\!\frac14\right)\Psi\left(\frac12\!\pm\!\sqrt{b_0}\right)
	-\!b_0\left(\frac32 b_0\!-\!\frac14\right).
	\end{eqnarray}

Since $\delta m_\chi^{2}(\varphi) = \lambda_{\chi}
\varphi_{0} \delta \varphi,\,\delta g^{\mu\nu} = {\rm diag}\,(0,
-2 g^{ab} \delta a/a)$, with $\delta\varphi$ and $\delta a$ given
by eqs.(\ref{deltavarphi}) and (\ref{deltacosm}), the final result for
a scalar field contribution to
$\delta{\mbox{\boldmath $\Gamma$}}_{\rm 1-loop}$ reads
up to terms $O(m_P^2/\varphi_{0}^{2})$
	\begin{eqnarray}
	&&\delta{\mbox{\boldmath $\Gamma$}}_{\rm 1-loop}^\chi
	= \frac{1}{2}\delta\zeta_\chi'(0)
	-\frac{1}{2}\delta\zeta_\chi(0)\ln\frac{H^{2}}{\mu^{2}},\nonumber\\
	&&\delta \zeta_\chi(0) =
	-\frac{9m_{P}^{2} |\xi| (1+\delta)
	\lambda_{\chi}^{2}}{\pi\lambda^{2}
	\varphi_{0}^{2}}\,\kappa,\,\,\,\,
	\delta \zeta_\chi'(0) = \frac{9m_{P}^{2} |\xi| (1+\delta)
	\lambda_{\chi}^{2}}{\pi\lambda^{2} \varphi_{0}^{2}}\,
	\kappa\,\ln\frac{6|\xi|\lambda_{\chi}}{\lambda},\label{deltazeta'}\\
	&&\kappa = \frac{\pi}{96}
	+\frac{\pi \ln 2}{12} - \frac{1}{72}
	\approx 0.2 .                                   \label{kappa}
	\end{eqnarray}

The leading terms of $\delta \zeta(0)$ and $\delta \zeta'(0)$ for
vector and spinor fields have a similar form and when composed
with the contribution of Higgs multiplets above
give rise to the total $\delta{\mbox{\boldmath $\Gamma$}}_{\rm 1-loop}$.
Similarly to the unperturbed part (\ref{one-loop-contr3}) it expresses
as a function of the universal combinations of coupling constants
${\mbox{\boldmath $A$}}$ and ${\mbox{\boldmath $B$}}$ given by
Eqs.(\ref{fieldcontent}) and (\ref{fieldcontent1})
	\begin{eqnarray}
	&&\delta{\mbox{\boldmath $\Gamma$}}_{\rm 1-loop} =
	\frac{9|\xi|(1+\delta)m_{P}^{2}
	}{\pi\lambda \varphi_{0}^{2}} \kappa
	{\mbox{\boldmath $A$}} \ln \frac{\varphi_{0}^{2}}{\mu^2}
	-\frac{9|\xi|(1+\delta)m_{P}^{2}
	}{\pi\lambda \varphi_{0}^{2}} \kappa
	{\mbox{\boldmath $B$}}.                \label{irregcontrib}
	\end{eqnarray}

\section{Probability maximum of the distribution function}
\hspace{\parindent}
Combining the tree-level part (\ref{actionexpansion})-(\ref{action1}) with
the contributions of the inflaton mode (\ref{inflatonmode}) and perturbative
contributions of the one-loop effective action (\ref{one-loop-contr3})
and (\ref{irregcontrib}) we finally arrive at the distribution function
(\ref{4.3.2}) of inflationary cosmologies
	\begin{eqnarray}
	&&\rho_{T,NB}(\varphi_0)=N\,\exp\frac{3|\xi|}{\lambda}
	\left[\,\mp\frac{m_P^2}{\varphi_0^2}\alpha_{\pm}
	-|\xi|{\mbox{\boldmath $A$}}
	\ln\frac{\varphi_0^2}{\mu^2}\right.  \nonumber\\
	&&\qquad\qquad\qquad\qquad\qquad\qquad\qquad\qquad\left.
	-\beta{\mbox{\boldmath $A$}}\frac{m_P^2}{\varphi_0^2}
	\ln\frac{\varphi_0^2}{\mu^2}
	+O\left(\frac{m_P^2}
	{|\xi|\varphi_0^2}\right)\,\right],  \label{rhofinal}
	\end{eqnarray}
where $N$ is a field independent normalization factor and $\alpha_{\pm}$ and
$\beta$ are the following functions of coupling constants of the model
	\begin{eqnarray}
	&&\alpha_{\pm}=8\pi(1+\delta)\mp\frac{1+2\delta}{4\pi}
	\left(\frac32 {\mbox{\boldmath $A$}}+{\mbox{\boldmath $B$}}\right)
	\pm 3\kappa{\mbox{\boldmath $B$}}
	\frac{1+\delta}\pi,                         \label{alpha}\\
	&&\beta=\frac{1+2\delta}{4\pi}
	-3\kappa\frac{1+\delta}\pi,                 \label{beta}
	\end{eqnarray}
involving the parameter $\delta$ (\ref{delta}) and universal combinations
${\mbox{\boldmath $A$}}$ and ${\mbox{\boldmath $B$}}$.

The second term in the exponential of (\ref{rhofinal}) confirms the
conclusions of \cite{BKam:norm,Bar-Rep,Bar-Kam-Scale} that loop corrections
can drastically change the predictions of classical theory and suppress
the contribution of the over-Planckian energy scales due to big positive
anomalous scaling of the theory
	\begin{eqnarray}
	\rho_{T,NB}(\varphi_0)\sim \varphi_0^{-Z-2},\,\,\,
	\varphi_0\rightarrow\infty,                       \label{sup}
	\end{eqnarray}
with $Z$ given by eq.(\ref{5.2.3}). Extra power $-2$ of $\varphi_0$ here
comes from the contribution of the inflaton mode (\ref{inflatonmode})
neglected in (\ref{rhofinal}) within the $O(m_P^2/\varphi_0^2)$
accuracy but important for $\varphi_0\rightarrow\infty$. For positive
$Z=6|\xi|^2{\mbox{\boldmath $A$}}/\lambda$ this
asymptotics can be regarded as a justification of a semiclassical
expansion.

The equation for the extremum of the obtained distribution at
$\varphi_0=\varphi_I$ can be represented in the form
	\begin{eqnarray}
	\pm\frac{\varphi_I^2}{m_P^2}=\frac{\alpha_{\pm}}{|\xi|
	{\mbox{\boldmath $A$}}}+\frac{\beta}{|\xi|}
	\left(\ln\frac{\varphi_I^2}{\mu^2}-1\right)
	+O\left(\frac1{|\xi|^2}\right),             \label{phiI}
	\end{eqnarray}
where plus or minus signs correspond to the tunnelling or no-boundary
wavefunctions respectively. To analyze the existence of its solution
we shall have to prescribe a certain reasonable range
of parameters and try solving it by iterations. In the next section we
shall briefly discuss the model of nonminimal inflation with big
$|\xi|\simeq 2\times 10^4\gg 1$ and $\delta=O(1)$ that was used in our
previous work \cite{Bar-Kam-Scale} as a good candidate for a quantum origin
of the early Universe at a sub-Planckian energy scale (around GUT scale).
In \cite{Bar-Kam-Scale} we only
took into account the tree-level part and the leading logarithmic behaviour
of the one-loop effective action, which correspond to retaining in
(\ref{rhofinal}) only the first two terms of the exponential with the
parameter $\alpha$ truncated to the first (${\mbox{\boldmath $A$}}$ and
${\mbox{\boldmath $B$}}$-independent) term of (\ref{alpha}).
In \cite{Bar-Kam-Scale} it was shown that the requirement of the minimal
admissible duration of the inflationary stage (\ref{N}) imposes upper
bound on the combination of coupling constants
${\mbox{\boldmath $A$}}\simeq 1.3$. Here we shall show that the qualitative
estimates of \cite{Bar-Kam-Scale} remain also true after the inclusion of
perturbative corrections obtained above.

The justification of the results of \cite{Bar-Kam-Scale} consists in the
observation that the second term in the right-hand side of (\ref{phiI})
can be regarded small and taken by perturbations. Indeed, from the upper bound
on ${\mbox{\boldmath $A$}}$ we can assume that both ${\mbox{\boldmath $A$}}$
and ${\mbox{\boldmath $B$}}$ are of the order of magnitude one
	\begin{eqnarray}
	{\mbox{\boldmath $A$}}=O(1),\,\,\,
	{\mbox{\boldmath $B$}}=O(1).             \label{O(1)}
	\end{eqnarray}
This follows from the comparison of expressions (\ref{fieldcontent}) and
(\ref{fieldcontent1})
and a natural assumption that the estimate for ${\mbox{\boldmath $B$}}$
follows from that for ${\mbox{\boldmath $A$}}$, unless strong cancellation
takes place between different separately big terms of the expression
(\ref{fieldcontent}) for ${\mbox{\boldmath $A$}}$. From these bounds we see
that $\alpha_{\pm}=8\pi(1+\delta)+O(1/10),\,\beta=O(1/10)$
due to the numerical values of the coefficients in
(\ref{alpha})-(\ref{beta}) and the fact that $\delta=O(1)$. Then it follows
that $\beta{\mbox{\boldmath $A$}}/\alpha_{\pm} \sim
{\mbox{\boldmath $A$}}/80\pi(1+\delta)\sim {\mbox{\boldmath $A$}}/250$
and, therefore, the second term in the right-hand side of (\ref{phiI})
can indeed be treated perturbatively with a good smallness parameter
${\mbox{\boldmath $A$}}/250$. With the same precision the expression for
$\alpha_{\pm}$ reduces
in the leading order to $\alpha=8\pi(1+\delta)$. It is needless to say that
for smaller values of ${\mbox{\boldmath $A$}}$ and ${\mbox{\boldmath $B$}}$
this approximation works even better, because the leading term in the
right-hand side of (\ref{phiI}) grows and the corrections decrease for
${\mbox{\boldmath $A$}}, {\mbox{\boldmath $B$}}\rightarrow 0$.

For $\delta>-1$ (which is the case of a standard classical scenario with a
finite duration of inflation, see Sect.2) the leading order solution of
(\ref{phiI}) exists only for the case of a tunneling wavefunction
(sign plus) and coincides with the result of \cite{Bar-Kam-Scale}. Thus,
the parameters of the inflation probability maximum -- the mean
value of the inflaton field $\varphi_I$ and its quantum dispersion
$\Delta\varphi=[-d^2 {\rm ln}\,{\mbox{\boldmath$\rho$}}(\varphi_I)/
d\varphi_I^2]^{-1/2}$ are
	\begin{eqnarray}
	\varphi_{I} = m_{P}\sqrt{\frac{8\pi(1+\delta)}{|\xi|
	{\mbox{\boldmath$A$}}}},\,\,\,\,\,
	\Delta\varphi=\frac{\varphi_I}
	{\sqrt{12{\mbox{\boldmath $A$}}}}
        \frac{\sqrt{\lambda}}{|\xi|}.              \label{peak}
	\end{eqnarray}
For the no-boundary quantum state of the Universe the peak can be
realized only for $\delta < -1$ and, thus corresponds to the
classical scenario with endless inflation stage \cite{Bar-Kam-Scale}.

\section{Nonminimal inflation and particle physics of the early
Universe}
\hspace{\parindent}
The present state of inflation theory is consistent with
observations of the
cosmic microwave background radiation anisotropy in
the COBE \cite{COBE} and Relikt \cite{Relikt} satellite experiments.
In the chaotic inflationary model with a nonminimal inflaton
field (\ref{4.2.3}) the spectrum of perturbations compatible
with these measurements can be acquired in the range of
coupling constants $\lambda/\xi^2\sim 10^{-10}$ \cite{SalopBB,Salopek}
(the experimental bound on the gauge-invariant \cite{BST} density
perturbation $P_{\zeta}(k)=N^2_k (\lambda/\xi^2)/8\pi^2$ in the $k$-th
mode "crossing" the horizon at the moment of the e-foldings number $N_k$).
The main advantage of this model is that it allows one to avoid an
unnaturally small value of $\lambda$ in the minimal inflaton model
\cite{Linde} and replace it with the GUT compatible value
$\lambda\simeq 0.05$, provided $\xi\!\simeq\! -2\!\times\! 10^4$ is chosen
to be related to the ratio of the Planck scale to a
typical GUT scale, $|\xi|\sim m_P/v$. For these values of coupling constants
the parameter (\ref{delta}) is $\delta\sim 8\pi v/m_P\sim 10^{-3}$ (thus
easily satisfying a much weaker upper bound $\delta=O(1)$ assumed above).
As far as it concerns the Hubble parameter (\ref{Hubble}) and the number
of e-foldings (\ref{N1}) at the inflationary peak (\ref{peak}) obtained
above, in the leading order in $|\xi|\gg 1$ they are given by
	\begin{eqnarray}
        &&H(\varphi_I)=
        m_{P}\frac{\sqrt{\lambda}}{|\xi|} \sqrt{\frac{2\pi(1+\delta)}
	{3{\mbox{\boldmath $A$}}^2}}, \label{7.1} \\
        &&{\mbox{\boldmath $N$}}(\varphi_I)=
        \frac{8\pi^2}{\mbox{\boldmath $A$}}        \label{4.3}
	\end{eqnarray}
and satisfy the bound  ${\mbox{\boldmath $N$}}(\varphi_I)\geq 60$
with a single restriction on ${\mbox{\boldmath $A$}}$, ${\mbox{\boldmath
$A$}}\leq1.3$. This restriction gives rise to the bounds (\ref{O(1)}) and,
therefore, justifies the smallness of the perturbative corrections
of the above type, that were neglected in our previous paper
\cite{Bar-Kam-Scale}. These perturbative corrections contained in the
second term of the equation (\ref{phiI}), unfortunately, violate the
conclusion of \cite{Bar-Kam-Scale} on the absence of renormalization
ambiguity in the energy scale of inflation, but as we see this ambiguity
in the choice of $\mu^2$ is strongly suppressed by the smallness of
the ratio $\beta{\mbox{\boldmath $A$}}/\alpha$.

On the other hand, the restriction on ${\mbox{\boldmath $A$}}$ justifies a
slow-roll approximation, because the corresponding smallness parameter
is $\dot\varphi/H\varphi\simeq-{\mbox{\boldmath $A$}}/96\pi^2
\sim-10^{-3}$. For the above small value of $\delta \ll1$ and
${\mbox{\boldmath $A$}}\simeq 1$,  the obtained
numerical parameters describe extremely sharp inflationary peak at
$\varphi_I$ with small width and sub-Planckian Hubble constant
	\begin{eqnarray}
	\varphi_I\simeq 0.03 m_P,\,\,
	\sigma\simeq 10^{-7}m_P,\,\,H(\varphi_I)\simeq 10^{-5}m_P,
	\end{eqnarray}
which is the most realistic range of the inflationary
scenario.  The smallness of the width does not, however, lead to its
quick quantum spreading: the commutator relations for operators
${\hat\varphi}$ and
$\dot{\hat\varphi}$, $[{\hat\varphi},\dot{\hat\varphi}]\simeq
i/(12\pi^2|\xi|a^3)$ \cite{SalopBB}, give rise at the beginning of
inflation, $a\simeq H^{-1}$, to a negligible
dispersion of
$\dot\varphi$, $\Delta\dot\varphi\simeq H^3/12\pi^2|\xi|\sigma
\simeq(8/{\mbox{\boldmath $A$}})(\sqrt{\lambda}/|\xi|)
|\dot\varphi|\ll|\dot\varphi|$. It is remarkable that the relative width
	\begin{eqnarray}
	\frac{\Delta\varphi}{\varphi_I}\sim\frac{\Delta H}H\sim 10^{-5}
	\end{eqnarray}
corresponds to the observable level of density perturbations, although it
is not clear whether this quantum dispersion
$\Delta\varphi$ is directly measurable now, because of the stochastic noise
of the same order of magnitude generated during the inflation and
superimposed upon $\Delta\varphi$.

All these conclusions are universal for a generic low-energy
model (\ref{partmodel}) and (apart from the choice
of $|\xi|$ and $\lambda$) universally depend on one parameter
${\mbox{\boldmath $A$}}$  of the particle physics model.  This quantity
should satisfy the lower and upper bounds
        \begin{eqnarray}
        \frac{\sqrt{\lambda}}{|\xi|}\ll
	{\mbox{\boldmath $A$}}\leq 1.3              \label{4.4}
        \end{eqnarray}
in order respectively to render $Z$ positive, thus suppressing over-Planckian
energy scales, and provide sufficient amount of inflation
(${\mbox{\boldmath $A$}}$ should not, certainly, be exceedingly close to
zero, not to suppress the dominant contribution of large $|\xi|$ in
(\ref{5.2.3})).

In \cite{Bar-Kam-Scale} the conclusion was made that this bound
suggests the quasi-supersymmetric nature of the
particle model, because supersymmetry can constrain the values of the
Higgs $\lambda_{\chi}$, vector gauge $g_A$ and Yukawa $f_{\psi}$
couplings so as to provide a subtle balance between the contributions of
bosons and fermions in (\ref{fieldcontent}) and fit the quantity
${\mbox{\boldmath $A$}}$ into a narrow range (\ref{4.4}). Now, however,
with the inclusion of corrections that go beyond the estimates of
\cite{Bar-Kam-Scale} we have also to provide the boundedness of
${\mbox{\boldmath $B$}}$ (\ref{fieldcontent1}) which is less obvious
to be compatible with the bound (\ref{4.4}), unless all the terms of
(\ref{fieldcontent}) and (\ref{fieldcontent1}) are separately small
due to small values of all coupling constants. In the latter case the
supersymmetry is not needed to explain the restrictions (\ref{4.4})
on the choice of a particle physics model. Still, supersymmetry
remains a reasonable conjecture consistent with this selection criterion
(\ref{4.4}) and sounds coherent with conclusions of \cite{Kam-Norm} where
supersymmetry was argued in the opposite case of small $|\xi|$.

\section{Conclusions}
\hspace{\parindent}
Thus, the same mechanism that suppresses the over-Planckian
energy scales also generates a narrow probability
peak in the distribution of tunnelling inflationary universes and
is likely to suggest the (quasi)supersymmetric nature of their particle
content. It seems to be consistent with microwave background
observations within the model with a strongly coupled nonminimal inflaton
field. A remarkable feature of this result is that it is mainly based on one
small parameter -- the dimensionless ratio of two major energy scales,
the GUT and Planck ones, given by the combination of the coupling
constants $\sqrt{\lambda}/|\xi|\simeq 10^{-5}$.

Big value of $|\xi|$ is actually responsible for the fact that the one-loop
corrections qualitatively change the tree-level behaviour and produce
the inflationary peak of the above type. In the absence of powerful
nonperturbative methods there is no rigorous proof that the inclusion of
multi-loop orders will not destroy this one-loop effect. However, the
following qualitative arguments support the conjecture that this will not
happen. Point is that the effective gravitational constant in this model is
inverse proportional to $m_P^2+8\pi|\xi|\varphi^2$
and, thus, large $|\xi|$ might improve the loop expansion \cite{BKK} by
suppressing the contribution of multi-loop orders. On the other hand,
the power-law mechanism (\ref{sup}) of suppressing the high-energy scales
is independent of the renormalization ambiguity (the parameter $\mu^2$).
This also gives a hope that the obtained effect is robust against inclusion
of multi-loop corrections. All these corrections are weighted by this
suppressing one-loop factor and are small at least in some range of $\varphi$ above the
probability peak value $\varphi_I$ where all their curvature invariants have
GUT values rather than Planckian ones. Therefore, irrespective of what
happens at Planckian scales this peak at GUT scale will be separated from
the unknown Planckian domain by the region with very small values of
the distribution function and, at least heuristically, will not
interfere with nonperturbative quantum gravity or its more fundamental
(stringy) generalization.

Obviously, the large value of $|\xi|$ at sub-Planckian (GUT) scale
requires explanation which might be based on the renormalization
group approach (and its extension to non-renormalizable theories
\cite{BKK}). As shown in \cite{BKK}, quantum gravity with nonminimal
scalar field has an asymptotically free conformally
invariant ($\xi=1/6$) phase at over-Planckian regime, which is
unstable at lower energies. It is plausible to conjecture that this
instability can lead (via composite states of the scalar field) to
the inversion of the sign of running $\xi$ and its growth at the GUT
scale, thus making possible the proposed inflation applications.

From the viewpoint of the theory of the early universe,
the obtained results give a strong preference to the
tunneling quantum state. Debates on advantages of
the tunneling versus no-boundary wavefunction has a long history
\cite{Linde,Vilenkin,rhoT,Rubak,Vilenkin:tun-HH}. At present,
in the cosmological context the tunnelling
proposal seems to be more useful and conceptually clearer than
the no-boundary one, because for its interpretation one should
not incorporate vague ideas of the third quantization of gravity
which inevitably arise in the no-boundary case: splitting the
Lorentzian wavefunction in positive and negative frequency
parts and separately calculating their probability distributions.
On the other hand, the formulation of the tunneling proposal is
not so aesthetically closed, for it involves imposing outgoing
condition after the potential barrier, the unit normalization
condition -- before the barrier at $a=0$, the requirement of
the normalizability in variables $f$ \cite{QGrav}, etc. And all this in
contradistinction with the closed path-integral formulation of
the no-boundary proposal, automatically providing many of the above
properties. On the other hand, outside of the cosmological
framework, in particular, within the scope of the wormhole and
black hole physics, the tunneling proposal seems to be helpless.
Moreover, at the overlap of the cosmological framework with the
theory of the virtual black holes it leads to contradictions
signifying that the quantum birth of bigger black holes is more
probable than of the small (Planckian) ones \cite{Bousso}.
All these arguments can hardly be conclusive, because it might
as well happen that the difference between the no-boundary and
tunnelling wavefunctions should be ascribed to the open problem
of the correct quantization of the conformal mode. Note that the
normalizability criterion for the distribution function  and its
algorithm (\ref{4.3.2}) do not extend to the low-energy limit
$\varphi\rightarrow 0$, where a naively computed no-boundary
distribution function blows up to infinity, the slow-roll
approximation becomes invalid, etc. This is a domain related to a
highly speculative (but, probably, inevitable) third quantization
of gravity \cite{Coleman}, which goes beyond the scope of this paper.
Fortunately, this domain is also separated from the obtained
inflationary peak by a vast desert with practically zero density
of the quantum distribution, which apparently
justifies our conclusions disregarding the ultra-infrared physics of
the Coleman theory of baby universes and cosmological constant
\cite{Coleman}.

\section*{Acknowledgements}
\hspace{\parindent}
The authors benefitted from helpful discussions with Don N.Page,
V.A.Rubakov, D.Salopek, A.A.Starobinsky and A.V.Toporensky.
This work was supported by the Russian Foundation for Basic Research
under grants 96-02-220, 96-02-16287 and 96-02-16295 and the European
Community Grant INTAS-93-493. Partly this work has been made possible
also due to the support by the Russian Research Project
``Cosmomicrophysics''.

\newpage
\section*{\centerline{Figure captions}}
{\bf Fig.1} Graphical representation of the Lorentzian spacetime
${\mbox{\boldmath $L$}}$ nucleating  at the bounce surface $\Sigma_{B}$ from
the Euclidean manifold ${\mbox{\boldmath $E$}}$ of the no-boundary type,
having the topology of the four-dimensional ball.
\\
\\
{\bf Fig.2} Two-dimensional minisuperspace of the scale factor $a$ and the
inflaton field $\phi$ in the chaotic inflation model. The Euclidean extremal
(in the slow roll approximation) starts at $a=0$ with large initial value
$\phi_0$ in the form of a trajectory that is reflected from the caustic
$\Sigma_B$, $a\simeq 1/H(\phi)$, and enters the region
$\phi\rightarrow\infty,\,\, a\rightarrow 0$.
\\
\\
{\bf Fig.3} Graphical representation of calculating the quantum
distribution of tunnelling Lorentzian universes: a composition of
the combined Euclidean-Lorentzian spacetime
${\mbox{\boldmath $M$}}_{-}\cup{\mbox{\boldmath $L$}}$
with its orientation reversed and complex conjugated copy
${\mbox{\boldmath $M$}}_{+}\cup{\mbox{\boldmath $L^*$}}$ results in the
doubled Euclidean manifold ${\bf 2}{\mbox{\boldmath $M$}}$ --
the gravitational instanton carrying the Euclidean effective action
of the theory. The cancellation of the Lorentzian domains
${\mbox{\boldmath $L$}}$ and ${\mbox{\boldmath $L^*$}}$ reflects
unitarity of the theory in the physical spacetime of Lorentzian signature.

\end{document}